\documentclass[twocolumn,superscriptaddress,showpacs,preprintnumbers,amsmath,amssymb]{revtex4-1}

\usepackage{amsmath}
\usepackage{latexsym}
\usepackage{graphicx}
\usepackage{dcolumn}

\newcommand{\bra}[1]{\langle #1 |}
\newcommand{\ket}[1]{| #1 \rangle}

\newcommand{\im}{\dot{\iota}\,}
\newcommand{\kz}{\mathsf{k}_z}
\newcommand{\kzi}[1]{\mathsf{k}_{z, #1}}
\newcommand{\kt}{\vec{k} \,}
\newcommand{\hz}{\hat{\mathbf{z}}}
\newcommand{\hkp}{\hat{\mathbf{s}}}
\newcommand{\hkb}{\hat{\mathbf{k}}_\bot}
\newcommand{\rv}{\mathbf{r}}
\newcommand{\kv}{\mathbf{k}}
\newcommand{\km}{\mathsf{k}}
\newcommand{\ev}{\mathbf{E}}
\newcommand{\bv}{\mathbf{B}}
\newcommand{\uv}{\mathbf{U}}
\newcommand{\vv}{\mathbf{V}}
\newcommand{\es}{\mathcal{E}}
\newcommand{\intd}{\int d \,}
\newcommand{\intdh}{\int^\prime d \,}
\newcommand{\pf}{{{\text{\tiny (+)}}}}
\newcommand{\nf}{{{\text{\tiny (--)}}}}
\newcommand{\va}{\hat{\mathbf{v}}_{\alpha}}

\newcommand{\vecP}{\hat{\boldsymbol{\rho}}}

\newcommand{\arctanh}{\text{arctanh}}
\begin{document}

\title{Cavity-QED of a leaky planar resonator coupled \\ to an atom and an input single-photon pulse}

\author{Denis Gon\c{t}a}
\email{denis.gonta@mpl.mpg.de}
\affiliation{Institute of Optics, Information and Photonics,
             Friedrich-Alexander-University Erlangen-Nuremberg,
             Staudtstrasse 7, 91058 Erlangen, Germany}

\author{Peter van Loock}
\email{loock@uni-mainz.de}
\affiliation{Institute of Physics,
             Johannes Gutenberg University Mainz,
             Staudingerweg 7, 55128 Mainz, Germany}

\date{\today}

\begin{abstract}
In contrast to the free-space evolution of an atom governed by a
multi-mode interaction with the surrounding electromagnetic vacuum,
the evolution of a cavity-QED system can be characterized by just
three parameters, (i) atom-cavity coupling strength $g$, (ii) cavity
relaxation rate $\kappa$, and  (iii) atomic decay rate into the
non-cavity modes $\gamma$. In the case of an atom inserted into a
planar resonator with an input beam coupled from the outside, it has
been shown by Koshino [Phys.~Rev.~A~\textbf{73},~053814~(2006)] that
these three parameters are determined not only by the atom and
cavity characteristics, but also by the spatial distribution of the
input pulse. By an \textit{ab-initio} treatment, we generalize the
framework of Koshino and determine the cavity-QED parameters of a
coupled system of atom, planar (leaky) resonator, and input
single-photon pulse as functions of the lateral profile of the pulse
and the length of resonator. We confirm that the atomic decay rate
can be suppressed by tailoring appropriately the lateral profile of
the pulse. Such an active suppression of atomic decay opens an
attractive route towards an efficient quantum memory for long-term
storage of an atomic qubit inside a planar resonator.
\end{abstract}


\maketitle

\section{Introduction}

Cavity quantum electrodynamics (cavity-QED) is a research field that studies
electromagnetic fields in confined spaces and radiative properties of atoms in
such fields. Experimentally, the simplest example of such a system is a single
atom interacting with a single mode of a high-finesse resonator \cite{rpp69}.
This system bears an excellent framework for quantum communication and information
processing, in which atoms and photons are interpreted as bits of quantum
information and their mutual interaction provides a controllable entanglement
mechanism \cite{rmp79}.

Remarkably, the evolution of a cavity-QED system can be well characterized by just
three parameters: (i) atom-cavity coupling strength  $g$, (ii) cavity relaxation
rate $\kappa$, and (iii) atomic decay rate into the non-cavity modes $\gamma$. The
cavity-QED effects become manifest clearly when the atom-cavity coupling $g$ is much
larger than the atomic decay rate $\gamma$ and the cavity relaxation rate $\kappa$, at
the same time. These two conditions define the (so-called) \textit{strong-coupling}
regime of atom-cavity interaction that ensures that the energy exchange between the
constituents is reversible and develops faster than losses due to the cavity relaxation
and the atomic decay. In the resonant regime, i.e., when the cavity resonant frequency
matches the atomic transition frequency, the reversibility of energy exchange ensures
that the coherent (unitary) part of atom-cavity evolution is governed by the
Jaynes-Cummings Hamiltonian \cite{jc}
\begin{equation}\label{jc}
H_\text{JC} = \hbar \, g \left( c \, \sigma^\dagger + c^\dagger \, \sigma \right) \, ,
\end{equation}
where $c$ and $c^\dag$ denote the cavity mode annihilation and creation operators, while
$\sigma$ and $\sigma^\dag$ are the atomic lowering and raising operators, respectively.
This Hamiltonian describes the interaction of a two-level atom with a single-mode light
field that is confined inside the resonator.

During the last decades, single-mode resonators with typically spherical
mirrors have been fabricated and utilized in various cavity-QED experiments. It was
demonstrated that resonators with spherical mirrors can operate in the strong-coupling
regime, such that the coherent part of the atom-cavity evolution is described by the
Hamiltonian (\ref{jc}) \cite{prl93}. Although a planar Fabry-Perot resonator with a
coupled atom is used to illustrate the main
features of cavity-QED, there is an essential difference between the resonator with
spherical mirrors used in typical cavity-QED experiments and a Fabry-Perot resonator
with two coplanar mirrors. Namely, even in the case of perfect lossless mirrors, a planar
(Fabry-Perot) resonator is intrinsically multimode with a spectrally dense continuum
of modes. Due to this essential difference, the cavity-QED parameters
$(g, \, \kappa, \, \gamma)$ cannot be identified straightforwardly in the case of
an atom coupled to a planar resonator.

In recent years, however, an impressive experimental progress has been achieved in
fabricating various planar like resonators, i.e., two-dimensional microwave circuits
(circuit-QED) \cite{pra75}, fiber-based (FFP) cavities \cite{njp12}, and diverse
micro-cavities \cite{kav}. Triggered by this experimental progress, the recent
\cite{pra72, prl100, prb78, pra82, pra86} and past \cite{pra35, pra43, DK, oc152, review,
pra60} theoretical developments devoted to planar cavities have acquired an increasing
attention. Although it is commonly agreed that the Rabi oscillations cannot occur in
an interacting system of an atom and a planar resonator because of a weak atom-cavity
coupling, it was pointed out in Refs.~\cite{pra51, K} that such system can still exhibit
Rabi oscillations, similar to those of a cavity-QED system, once the planar resonator
is excited by a coherent external beam. In other words, provided that a light pulse
penetrates the resonator from outside with an appropriately tailored spatial distribution,
the coupling strength of an (otherwise weakly interacting) atom-cavity system can be
dramatically enhanced, leading to Rabi oscillations.

The experimental evidences which support the existence of Rabi oscillations in a
coupled exciton-photon system confined in a planar microcavity and exposed to an
external coherent beam has been presented in Refs.~\cite{prb50, prl69}. Since the
coherent part of both evolutions associated with confined exciton-photon and
atom-photon coupled systems is governed by the Jaynes-Cummings Hamiltonian
(\ref{jc}), these experiments provide compelling arguments that an interacting
system of three constituents, i.e., (i) an atom weakly coupled to (ii) a planar
resonator, and (iii) a spatially tailored input pulse, is capable to reproduce
the cavity-QED evolution. Similar to the cavity-QED system, furthermore, this
(atom-cavity-pulse) system is characterized by a set of parameters determined not
only by the atom, cavity, and reservoir characteristics, but also by the spatial
distribution of the input pulse. To our best knowledge, the problem of identifying
these parameters has been addressed solely by Koshino in Ref.~\cite{K}.

Using the (so-called) form-factor formalism, in this reference, the
author suggested three functions which correspond to the cavity-QED
triplet $(g, \, \kappa, \, \gamma)$, and he showed their dependence
on the spatial distribution of the input pulse. As a consequence of
the developed formalism, it was demonstrated how to suppress the
atomic decay $\gamma$ by tailoring appropriately the spatial distribution
of this input pulse. However, Koshino introduced four simplifying
assumptions in his framework, namely, (i) the evolution of the
coupled atom-cavity-pulse system was described by an \textit{ad hoc}
Hamiltonian, (ii) the light field had only one (fixed) polarization,
(iii) the atom was described by an averaged (in space) dipole, while
(iv) the planar resonator accommodated only one atomic wavelength.

In contrast to Koshino's approach, in this paper, we develop an
\textit{ab-initio} theoretical framework, in which we completely
exclude the above simplifications. In this generalized framework,
we derive the cavity-QED parameters of a coupled atom-cavity-pulse
system and reveal the dependence of these parameters on the
atom-cavity-reservoir characteristics and spatial distribution of
the input pulse. We find explicitly the optimal lateral profile that
yields a complete vanishing of the atomic decay rate and, thus, we
also find that the atomic decay can be efficiently suppressed by
coupling of an appropriate pulse to the resonator. Besides this
optimal pulse, we consider the situation in which a Hermite-Gaussian
beam penetrates the resonator from outside. We calculate cavity-QED
parameters for this case and reveal their dependence on the beam
waist and the cavity length.

The controllable suppression of the atomic decay opens an attractive
route towards an efficient quantum memory for long-term storage of a
single qubit that is encoded by a two-level atom coupled to the planar
resonator and an input pulse, while the atomic decay constitutes the
main source of decoherence. Such a quantum memory poses an essential
prerequisite for quantum information processing and quantum networking
applications like quantum repeaters \cite{rep} and quantum key
distribution \cite{qkd}. In this paper, however, we address solely the
physical aspects of the suggested quantum memory, i.e., the cavity
QED behavior of the coupled atom-cavity-pulse system, while a quantitative
characterization of the suggested quantum memory shall be addressed
in our future works.

The paper is organized as follows. In the next section, we analyze a
leaky planar resonator and derive the quantized electromagnetic
field produced inside and outside the resonator. In Sec.~II.C we
discuss the limit of perfect reflectivity, which is relaxed in
Sec.~II.D to the case of a high but finite reflectivity. Using the
total Hamiltonian of a coupled atom-cavity-pulse system derived in
Secs.~III.C and III.D, we introduce the form-factor formalism and
identify the cavity-QED parameters in Sec.~III.C. In Sec.~IV.A, we
evaluate these parameters by considering an optimal lateral profile
that yields a suppression of atomic decay, while an (experimentally
feasible) Hermite-Gaussian beam is considered in Sec.~IV.B. A
summary and outlook are given in Sec.~V.

\section{One-sided leaky cavity with planar geometry}

\begin{figure}
\begin{center}
\includegraphics[width=0.425\textwidth]{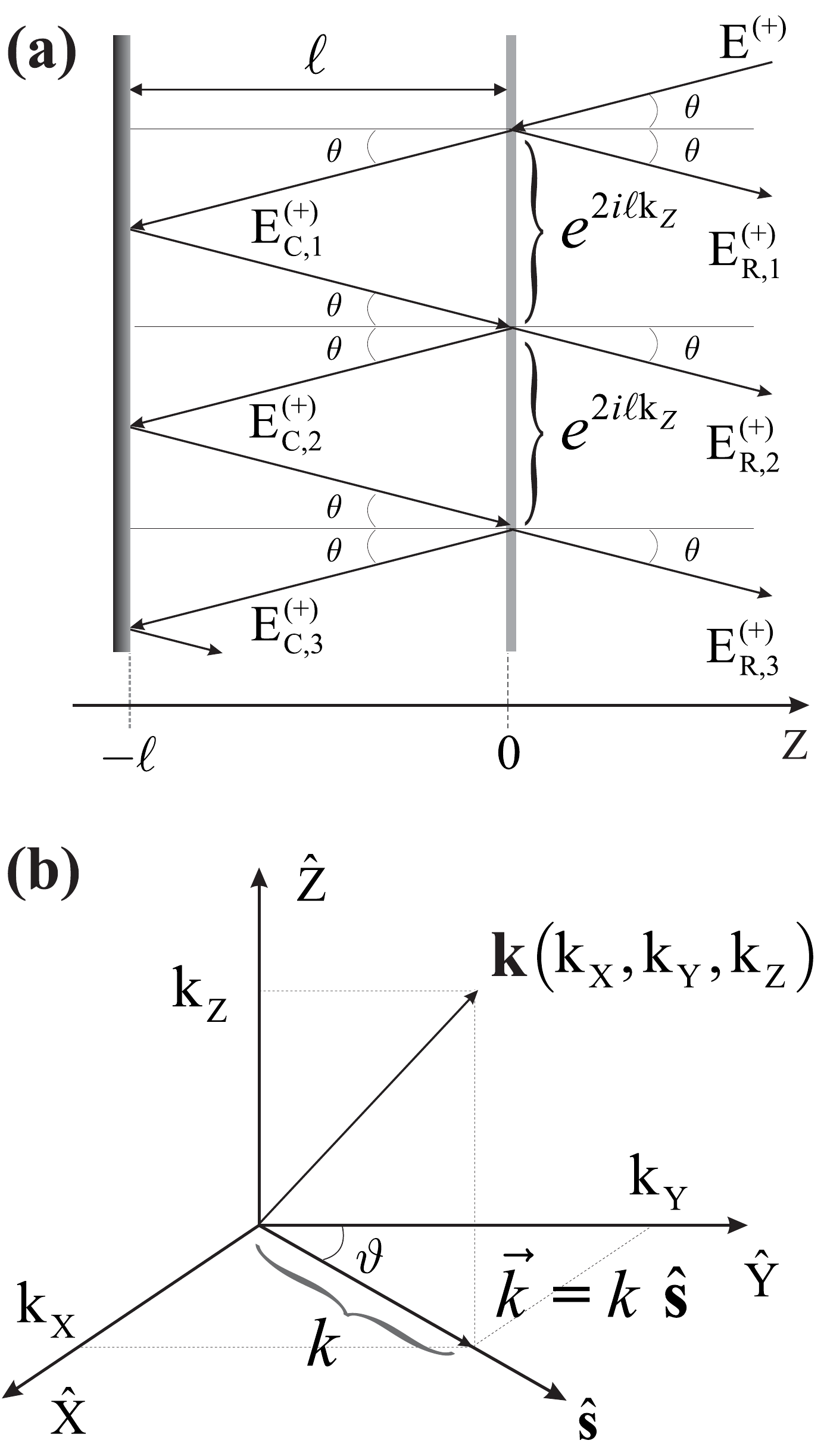} \\
\caption{(Color online) (a) Multiple-reflections method \cite{DK}
for an incident plane wave that penetrates the planar resonator from
the outside. (b) Cylindrical coordinate system in the reciprocal
space. See text for details.}
\label{fig1}
\end{center}
\end{figure}

In order to describe a two-level atom coupled to a field confined 
in a planar resonator, we have to consider first an empty resonator 
and determine the respective quantized electromagnetic field. In this 
section, we analyze the one-sided leaky cavity with planar geometry as 
shown in Fig.~\ref{fig1}(a). This cavity consists of a perfectly 
reflecting (solid) plane mirror located at $z = - \ell$ and a leaky 
(semitransparent) plane mirror located at $z = 0$.

As we mentioned in the introduction, there is an essential
difference between a resonator with spherical mirrors, used in
typical cavity-QED experiments, and a planar resonator. In the
latter confinement configuration, only the normal component of wave
vector (along the $\kz$-axis) can take discrete values inside a
perfectly reflecting (lossless) planar cavity, while the other two
components propagate freely. In a leaky planar resonator, in
contrast, the semitransparent mirror at $z=0$ causes the cavity
relaxation, i.e., the leakage of cavity photons and, therefore, even
the $z$-component of wave vector can never become completely
discrete. In contrast to the Koshino's treatment, the cavity
relaxation in our approach is not a predefined function. Instead, it
is determined by the transmissivity and reflectivity parameters of
planar resonator. This enables us to include both the intra-cavity
field and the field that leaks outside (or penetrates into the
resonator) in the same framework.

\subsection{Semitransparent dielectric-slab mirror}

Following the conventional approach (see Sec.~5.C in Ref.~\cite{SC}),
we model the semitransparent mirror by an idealized (infinitesimally)
thin layer of dielectric material, the so-called dielectric slab,
with the dielectric constant around $z=0$ given by
\begin{equation}
\epsilon(z) = \epsilon \left[ 1 + \eta \, \delta(z) \right] \, ,
\end{equation}
where $\epsilon$ denotes the permittivity of vacuum and $\eta$ is the positive and real
parameter that encodes the transparency (see below). In Ref.~\cite{DK}, Dutra and Knight
showed that the transmissivity and reflectivity of such a thin dielectric slab are given
by the expressions
\begin{subequations}\label{rt-semi}
\begin{eqnarray}
&& T_\bot (\kv) = \frac{2 \, \kz}
                     {2 \, \kz - \im \kv^2 \, \eta} \, ; \quad
T_\parallel (\kv) = \frac{2}{2 - \im \kz \, \eta} \, , \quad \\
&& R_\bot (\kv) = \frac{\im \kv^2 \, \eta}
                     {2 \, \kz - \im \kv^2 \, \eta} \, ; \quad
R_\parallel (\kv) = \frac{\im \kz \, \eta}
                    {\im \, \kz \, \eta - 2} \, , \quad
\end{eqnarray}
\end{subequations}
which fulfill the equalities
\begin{subequations}\label{cond0}
\begin{eqnarray}
&& \qquad | R_\alpha (\kv)|^2 + | T_\alpha (\kv)|^2 = 1  \, , \\
&& R_\alpha(\kv)^* \, T_\alpha(\kv) + T_\alpha(\kv)^* \, R_\alpha(\kv) = 0 \, ,
\end{eqnarray}
\end{subequations}
where $\kv$ is the wave vector, $\alpha = \bot$ refers to the component normal
to the plane of mirror (along the $z$ axis), and $\alpha = \parallel$ refers to
the component lying on the plane of mirror ($x-y$ plane).

Using Eqs.~(\ref{rt-semi}), one can readily check that
the mirror becomes completely transparent in the limit $\eta \rightarrow 0$,
while it becomes a perfect reflector in the limit $\eta \rightarrow \infty$,
i.e.,
\begin{equation}\label{rt-limit}
T_\alpha (\kv) =
\begin{cases}
1, \; \eta \rightarrow 0, \\
0, \; \eta \rightarrow \infty,
\end{cases}
R_\alpha (\kv) =
\begin{cases}
0, \; \eta \rightarrow 0, \\
m_\alpha, \; \eta \rightarrow \infty,
\end{cases}
\end{equation}
where $m_\bot = -1$ and $m_\parallel = 1$. The parameter $\eta$, therefore,
determines alone the transmissivity and reflectivity of the (leaky) mirror.

In our scheme, the leaky mirror at $z=0$ ensures also that a light pulse can penetrate the
resonator from outside. The leaky mirror, therefore, is supposed to have a high but non-perfect
reflectivity ($\eta \gg 1$) which, in this paper, is understood as a small deviation from
the perfect reflectivity limit. Since $T_\alpha (\kv)$ and
$R_\alpha (\kv)$ are constant in the perfect reflectivity limit [see Eqs.~(\ref{rt-limit})],
we can treat (to a good approximation) the transmissivity and reflectivity of a leaky mirror
as complex valued constants
\begin{subequations}
\begin{eqnarray}
&& \qquad T_\alpha (\kv) \cong T_\alpha \quad \text{and} \quad R_\alpha (\kv) \cong R_\alpha \, ; \label{rt-semi1} \\
&& |R_\alpha|^2 + | T_\alpha|^2 = 1  \, , \quad R_\alpha^* \, T_\alpha + T_\alpha^* \, R_\alpha = 0 \, .
\label{cond1}
\end{eqnarray}
\end{subequations}
On top of this, moreover, the expressions (\ref{rt-semi}) imply
\begin{equation}\label{rel}
\text{Re} \left( T_\alpha \right) \ll \text{Im} \left( T_\alpha \right) \quad \text{and} \quad
\text{Re} \left( R_\alpha \right) \gg \text{Im} \left( R_\alpha \right) \, .
\end{equation}

The assumption (\ref{rt-semi1}) together with relations (\ref{cond1}) and (\ref{rel})
suggest that $T_\alpha$ and $R_\alpha$ can be chosen in the form
\begin{equation}\label{new}
R_\alpha = m_\alpha \sqrt{1 - \tau^2} \, ; \quad T_\alpha = \im \, \tau \, , \quad
\tau \ll 1 \,
\end{equation}
in order to describe a leaky mirror that deviates only slightly from the perfect one.
This presentation implies that the limit of perfect reflectivity is reproduced up to
the first order of $\tau$, i.e., $R_\alpha$ starts to deviate from $m_\alpha$ to the
second order of $\tau$. Throughout this paper, therefore, we consider the expressions
(\ref{new}) to describe a leaky mirror, while the expressions (\ref{rt-semi}) are
considered to describe an arbitrarily semitransparent mirror or a perfectly 
reflecting mirror.

\subsection{One-sided planar resonator with a semitransparent mirror}

An unconfined light propagates in free space, such that the positive-frequency part 
of its electric field $\ev(\rv,t)$ is expressed as follows \cite{CT}
\begin{eqnarray}\label{elf-s}
\ev^\pf(\rv,t) &=&  \sum_\alpha \intd \kv \, \va(\kv) \, \es^\pf_\alpha(\kv) \,
e^{\im (\kv \cdot \rv - \km c \, t)} \notag \\
&\equiv& \intd \kv \, \ev^\pf (\kv, \rv) \, e^{-\im \km c \, t} \, , \label{pf1}
\end{eqnarray}
where $\km \equiv |\kv|$ denotes the modulus of wave vector, $\va(\kv)$ denotes 
the unit vector specifying the direction of a given electric-field component, 
while $\es^\pf_\alpha(\kv)$ denote the electric-field amplitude. We calculate how 
the one-sided planar resonator with a semitransparent 
mirror modifies the plane waves encoded by the expression $\ev^\pf (\kv, \rv)$. 
Having this modified expression, we then insert it back into Eq.~(\ref{pf1}) along 
with the quantum counterparts of the field amplitudes $\es^\pf_\alpha(\kv)$ and 
determine the quantized electric field in the presence of the resonator. With the 
help of quantized electric and magnetic fields, furthermore, we compute the total
electromagnetic energy in the physical space that includes regions inside and 
outside the cavity along with the region occupied by the leaky mirror.

Similar to the theory of a Fabry-Perot resonator \cite{BW}, we apply
the multiple-reflection approach by summing the reflected and
transmitted plane waves as depicted in Fig.~\ref{fig1}(a). We recall
that the solid mirror is a perfectly conducting plane that implies
the transformation of the amplitudes of the electric field
\begin{equation}\label{f-rt-solid}
\es^\pf_{\alpha}(\kv) \rightarrow m_\alpha \, \es^\pf_{\alpha}(\kv);
\quad m_\bot = -1, \quad m_\parallel = 1,
\end{equation}
In contrast to the perfect mirror at $z = -\ell$, the action of a
semitransparent mirror on the incident plane waves is determined by
the reflectivity and transmissivity (\ref{rt-semi}), which imply the
respective transformations of the amplitudes of the electric field
\begin{subequations}\label{f-rt-semi}
\begin{eqnarray}
&& \es^\pf_{\alpha}(\kv) \rightarrow R_\alpha (\kv) \, \es^\pf_{\alpha}(\kv) \, ; \\
&& \es^\pf_{\alpha}(\kv) \rightarrow T_\alpha (\kv) \,
\es^\pf_{\alpha}(\kv) \, .
\end{eqnarray}
\end{subequations}

Using the multiple-reflection approach with relations
(\ref{f-rt-solid}) and (\ref{f-rt-semi}), we compute the electric field inside
($-\ell \leq z < 0$) and outside ($z > 0$) the planar cavity region,
\begin{eqnarray}
&& \ev^\pf_\text{C}(\kv, \rv) = \sum_{i=1}^\infty \ev^\pf_{\text{C}, i} (\kv, \rv) \notag \\
                            &&= 2 \left[ \es^\pf_\parallel(\kv) \, L_\parallel(\kv) \, \cos[\kz (z + \ell)] \,
                                                    \frac{k}{\km} \, \hz \right. \notag \\
                                                    &&- \im \, \es^\pf_\parallel(\kv) \, L_\parallel (\kv) \, \sin[\kz (z + \ell)]
                                                        \frac{\kz}{\km} \, \hkp \notag \\
                                                        &&- \left. \im \, \es^\pf_\bot(\kv) \, L_\bot(\kv) \, \sin[\kz (z + \ell)] \, \hkb \right]
                                                        e^{\im \kt \cdot \rv}, \label{efc}
\end{eqnarray}
\begin{eqnarray}
&& \ev^\pf_\text{O}(\kv, \rv) = \ev^\pf (\kv, \rv) + \sum_{i=1}^\infty \ev^\pf_{\text{R},i} (\kv, \rv) \notag \\
                                                    &&= \left[ \es^\pf_\parallel(\kv) \left( e^{-\im \kz z} +
                                                        P_\parallel(\kv) \, e^{\im \kz z} \right) \, \frac{k}{\km} \, \hz \right. \notag \\
                                                        &&+ \es^\pf_\parallel(\kv) \left( e^{-\im \kz z} -
                                                        P_\parallel(\kv) \, e^{\im \kz z} \right) \frac{\kz}{\km} \, \hkp \notag \\
                                                        &&+ \left. \es^\pf_\bot(\kv) \left( e^{-\im \kz z} +
                                                        P_\bot(\kv) \, e^{\im \kz z} \right) \hkb \right] e^{\im \kt \cdot \rv}, \label{efo}
\end{eqnarray}
where $\kv = \{ \kz, k, \vartheta \}$ has been expressed in the
cylindrical coordinate basis, while $\kt = k \, \hkp$ is the (in-plane) 
wave vector lying on the plane of mirror as shown in Fig.~\ref{fig1}(b).
The orthogonal unit vectors $\hz$, $\hkp$, and $\hkb \equiv \hkp \times \hz$
determine the polarization of the resulting electric field, while
\begin{subequations}\label{lp-def}
\begin{eqnarray}
&& \qquad L_\alpha(\kv) \equiv \frac{T_\alpha(\kv)}{1 - e^{2 \, \im \ell \, \kz} \, m_\alpha \, R_\alpha(\kv)} \, ,
\label{l-def} \\
&& P_\alpha(\kv) \equiv R_\alpha(\kv) + T_\alpha(\kv) \, L_\alpha(\kv) \, m_\alpha \, e^{2 \, \im \ell \, \kz}
\label{p-def}
\end{eqnarray}
\end{subequations}
characterize the spectral response of resonator \cite{D}.

At this point, we introduce the quantum counterpart of the (positive-frequency)
field amplitude
\begin{equation}\label{ampl}
\es^\pf_\alpha(\kv) = \sqrt{\frac{2 \, \hbar \, \omega_\kv}{\epsilon (2 \pi)^3}} \, a_\alpha(\kv) \, ,
\end{equation}
where $\omega_\kv \equiv c \, |\kv| = c \, \km$, while $a_\alpha(\kv)$ is the 
photon annihilation operator that satisfies
\begin{subequations}\label{comm}
\begin{eqnarray}
&& [ a_\alpha(\kv), a^\dagger_{\alpha^\prime}(\kv^\prime) ]
        = \delta_{\alpha, \alpha^\prime} \, \delta \left( \kv - \kv^\prime \right),  \\
&& \hspace{1cm} [ a_\alpha(\kv), a_{\alpha^\prime}(\kv^\prime) ] = 0 \, .
\end{eqnarray}
\end{subequations}
In contrast to the free-space case, the above amplitude contains an extra factor 
of $2$ due to the perfect mirror restricting the field to the half-space only \cite{DK}.
By inserting Eqs.~(\ref{efc}) and (\ref{efo}) along with the amplitude (\ref{ampl}) 
into the integral (\ref{pf1}), we obtain the quantized electric field inside  
$\ev^\pf_\text{C}(\rv,t)$ and outside the cavity $\ev^\pf_\text{O}(\rv,t)$ region,
\begin{eqnarray}\label{elf}
&& \ev^\pf_\bullet(\rv,t) = \\
                    && = \sum_\alpha \intdh \kv \sqrt{\frac{2 \, \hbar \, \omega_\kv}{\epsilon (2 \pi)^3}} \,
                    \uv_{\alpha, \bullet}(\kv, z) \, a_\alpha(\kv)  \, e^{\im (\kt \cdot \rv - \omega_\kv t)} , \notag
\end{eqnarray}
where the integration (prime symbol) is restricted to the positive $\kz$,
while the subscript $\bullet$ denote C or O depending on the region inside 
or outside the cavity, respectively. In this expression,
\begin{subequations}\label{u-modes}
\begin{eqnarray}
\uv_{\bot, \text{C}}(\kv, z) &=& - 2 \, \im \, L_\bot(\kv) \, \sin[\kz (z + \ell)] \, \hkb \, ; \\
\uv_{\bot, \text{O}}(\kv, z) &=& \left( e^{-\im \kz z} + P_\bot(\kv) \, e^{\im \kz z} \right) \hkb \, ; \\
\uv_{\parallel, \text{C}}(\kv, z) &=& 2 \, L_\parallel(\kv) \, \left(
\cos[\kz (z + \ell)] \, \frac{k}{\km} \, \hz \right. \notag \\
                                  &-& \left. \im \, \sin[\kz (z + \ell)] \frac{\kz}{\km} \, \hkp \right) \, ; \\
\uv_{\parallel, \text{O}}(\kv, z) &=& \left( e^{-\im \kz z} +                                                                             P_\parallel(\kv) \, e^{\im \kz z} \right) \frac{k}{\km} \, \hz \notag \\
          &+& \left( e^{-\im \kz z} - P_\parallel(\kv) \, e^{\im \kz z} \right)
          \frac{\kz}{\km} \, \hkp \, . \quad
\end{eqnarray}
\end{subequations}
Moreover, the global mode functions
\begin{equation}
\uv_\alpha(\kv, \rv) \equiv e^{\im \kt \cdot \rv} \left[ \uv_{\alpha, \text{O}}(\kv, z) +
                                                         \uv_{\alpha, \text{C}}(\kv, z) \right]
\end{equation}
form an orthonormal set
\begin{equation}\label{orth-set}
\int d \rv \, \uv_\alpha(\kv, \rv) \cdot \uv^*_{\alpha^\prime}(\kv^\prime, \rv)
         = (2 \pi)^3 \, \delta_{\alpha, \alpha^\prime} \, \delta \left( \kv - \kv^\prime \right) \, ,
\end{equation}
where the integration over the $z$-axis is restricted to $-\ell < z < \infty$.
Using the electric field (\ref{elf}), we readily obtain the magnetic field inside
and outside the cavity region,
\begin{eqnarray}\label{mgf}
&& \bv^\pf_\bullet(\rv,t) = \\
                    && = \sum_\alpha \intdh \kv \sqrt{\frac{2 \, \hbar \, \omega_\kv}{c \, \epsilon (2 \pi)^3}}
                    \vv_{\alpha,\bullet}(\kv, z) \, a_\alpha(\kv)  \, e^{\im (\kt \cdot \rv - \omega_\kv t)} , \notag
\end{eqnarray}
where
\begin{equation}
\vv_{\alpha,\bullet}(\kv, z) = \frac{1}{\km} \left( \kt - \im \hz \frac{\partial}{\partial z}
                                              \right) \times \uv_{\alpha,\bullet}(\kv, z) \, .
\end{equation}

Owing to the quantized electric (\ref{elf}) and magnetic (\ref{mgf}) fields,
we compute the energy of the electromagnetic field in the space that includes
the regions inside and outside the cavity together with the region occupied by
the leaky mirror. This energy is given by the free-field Hamiltonian
\begin{eqnarray}\label{ham-t0}
H_\text{F} &=& \int_{-\infty}^\infty dx  \, dy
               \left[ \int_{-\ell}^0 dz \, \mathcal{H}_\text{C}(\rv,t) \right. \notag \\
           &+& \left.  \int_0^\infty dz \, \mathcal{H}_\text{O}(\rv,t) +
               \lim_{\xi \rightarrow \, 0} \int_{- \xi}^\xi dz \,
               \mathcal{H}_\text{L}(\rv,t) \right] \, , \quad
\end{eqnarray}
where $\mathcal{H}_\text{C}(\rv,t)$ and $\mathcal{H}_\text{O}(\rv,t)$ are the
energy densities of the electromagnetic fields inside and outside the cavity region,
respectively, while $\mathcal{H}_\text{L}(\rv,t)$ is the energy density associated
with the leaky mirror. Since the electric field vanishes inside a perfectly
reflecting mirror, there is no extra contribution to (\ref{ham-t0}). By following
the approach of Ref.~\cite{D}, it can be shown that the above Hamiltonian takes
the expected form
\begin{equation}\label{ham-t}
H_\text{F} = \sum_\alpha \intdh \kv \, \frac{\hbar \, \omega_\kv}{2} \left[ a_\alpha^\dagger(\kv) \,
            a_\alpha(\kv) + a_\alpha(\kv) \, a_\alpha^\dagger(\kv) \right]
\end{equation}
and describes the energy of an infinite set of harmonic oscillators
each characterized by the frequency $\omega_\kv$.

\subsection{The limit of perfect reflectivity}

In the previous section, we derived the quantized electric and magnetic fields in
the presence of a one-sided planar resonator as displayed in Fig.~\ref{fig1}(a) 
with an arbitrarily semitransparent mirror. We found that
the energy of the electromagnetic field, localized inside and outside the cavity and
inside the semitransparent mirror, is given by the Hamiltonian (\ref{ham-t}).
In this section, we reconsider the obtained results in the (lossless) limit of perfect
reflectivity, i.e., when the leaky mirror at $z = 0$ becomes a perfectly reflecting
mirror ($\eta \rightarrow \infty$). In this case, two types of photon field operators,
acting in the intra-cavity region and outside region (reservoir) emerge from
the global photon operator $a_\alpha(\kv)$. In the next sections, we consider these
operators for the case of a leaky cavity given by the condition $\tau \ll 1$ and
understood as a small deviation from the perfect-reflectivity case.

In order to proceed, we express Eq.~(\ref{elf}) in the form
\begin{eqnarray}\label{elf-c}
\ev^\pf_\text{C}(\rv,t) &=& \sum_\alpha \intdh \kv \sqrt{\frac{2 \, \ell \, \hbar \; \omega_\kv}
{\pi \, \epsilon \, (2 \pi)^3}} \times \\
&& \qquad \widetilde{\uv}_{\alpha, \text{C}} (\kv , z) \, |L_\alpha(\kv)|^2 \,
               \tilde{a}_\alpha(\kv)  \, e^{\im (\kt \cdot \rv - \omega_\kv \, t)} \, , \notag
\end{eqnarray}
where we introduced
$\uv_{\alpha, \text{C}}(\kv, z) \equiv L_\alpha(\kv) \, \widetilde{\uv}_{\alpha, \text{C}} (\kv, z)$
and
\begin{equation}
a_\alpha(\kv) \equiv \sqrt{\ell / \pi} \, L^*_\alpha(\kv) \, \tilde{a}_\alpha(\kv) \, . \label{def1}
\end{equation}
In the perfect-reflectivity limit, it was showed by Dutra and Knight in Ref.~\cite{DK} that
\begin{equation}
\lim_{\eta \rightarrow \infty} |L_\alpha(\kv)|^2
   = \frac{\pi}{\ell} \sum_{n = -\infty}^\infty \delta \left( \kz - \kzi{n} \right),
\end{equation}
where $\kzi{n} \equiv n \, \pi / \ell$. By inserting this expression into
Eq.~(\ref{elf-c}) and integrating over $\kz$, we obtain the electric field
inside the cavity in the limit of perfect reflectivity,
\begin{eqnarray}\label{elf-cl}
\ev^\pf_\text{CL}(\rv,t) &=&
\sum_{\alpha, n} \int d \kt \, \sqrt{\frac{\hbar \; \omega_{n, k}}{\epsilon (2 \pi)^2 \, \ell}} \times \\
&& \widetilde{\uv}_{\alpha, \text{C}} (\kzi{n}, \kt, z) \,
          \tilde{a}_\alpha(\kzi{n}, \kt) \, e^{\im ( \kt \cdot \rv - \omega_{n, k} t )}, \notag
\end{eqnarray}
where $\omega_{n, k} \equiv c \sqrt{ \kzi{n}^2 + k^{\, 2} }$ is the (quasi-mode)
frequency, and where the reduced mode functions $\widetilde{\uv}_{\alpha, \text{C}}$
form an orthogonal set
\begin{equation}\label{orth-set1}
\int dz \, \widetilde{\uv}_{\alpha, \text{C}}
(\kzi{n}, \kt, z) \cdot \widetilde{\uv}^*_{\alpha^\prime, \text{C}}
(\kzi{n^\prime}, \kt, z) = 2 \, \ell \; \delta_{n, n^\prime} \, \delta_{\alpha, \alpha^\prime} \,
\end{equation}
with the integration being restricted to $-\ell < z < 0$.

The photon annihilation operator $\tilde{a}_\alpha(\kzi{n}, \kt)$ in Eq.~(\ref{elf-cl})
depends on the discrete values of $n$ and continuous values of $\kt$.
Although the expression $\ev^\pf_\text{CL}(\rv,t)$ is formally identical to the
(positive-frequency parts of) electric field inside a lossless planar resonator
(see Ref.~\cite{DK}), we remind that the operator $\tilde{a}_\alpha(\kzi{n}, \kt)$
has been obtained from $a_\alpha(\kv)$ using the definition (\ref{def1})
and the discretization of $\kz$ component. In order to define the cavity
operator acting inside the cavity region only, we keep Eq.~(\ref{elf-cl}) with
replacing $\tilde{a}_\alpha(\kzi{n}, \kt)$ by operator $c_{\alpha,n}(\kt)$,
\begin{eqnarray}\label{elf-cp}
\ev^\pf_\text{CP}(\rv,t) &=&
\sum_{\alpha, n} \int d \kt \, \sqrt{\frac{\hbar \; \omega_{n,k}}{\epsilon (2 \pi)^2 \, \ell}}
                           \times \\
&& \qquad \widetilde{\uv}_{\alpha, \text{C}} (\kzi{n}, \kt, z) \,
          c_{\alpha, n}(\kt) \, e^{\im ( \kt \cdot \rv - \omega_{n,k} t )} \, , \notag
\end{eqnarray}
where we interpret $c_{\alpha,n}(\kt)$ as the cavity photon annihilation operator
that satisfies
\begin{subequations}\label{comm1}
\begin{eqnarray}
&& [ c_{\alpha,n}(\kt), c^\dagger_{\alpha^\prime, n^\prime}(\kt^\prime) ]
        = \delta_{\alpha, \alpha^\prime} \, \delta_{n, n^\prime} \,
          \delta \left( \kt - \kt^\prime \right),  \label{comm1a} \\
&& \hspace{1.5cm} [ c_{\alpha,n}(\kt), c_{\alpha^\prime, n^\prime}(\kt^\prime) ] = 0 \, , \label{comm1b}
\end{eqnarray}
\end{subequations}
and is characterized by the quasi-mode frequency $\omega_{n,k}$.

In order to derive $c_{\alpha,n}(\kt)$ in terms of global operator
$a_\alpha(\kv)$, we multiply (scalarly) both Eqs.~(\ref{elf})
and (\ref{elf-cp}) by $\widetilde{\uv}^*_{\alpha, \text{C}} (\kzi{n^\prime}, \kt, z)$
and integrate them over $z$ from $- \ell$ to $0$ using the property (\ref{orth-set1}).
We equate the resulting expressions and solve them for the operator $c_{\alpha,n}(\kt)$,
\begin{eqnarray}\label{c-op}
&& c_{\alpha,n}(\kt) = \frac{1}{2 \,\sqrt{\pi \, \ell}} \intdh \kz \sqrt{\frac{\omega_\kv}
                       {\omega_{n,k}}} \, L_\alpha (\kv) \, a_\alpha (\kv) \times \\
                    && \qquad e^{\im ( \omega_{n,k} - \omega_\kv ) \, t}
                       \int dz \; \widetilde{\uv}_{\alpha, \text{C}} (\kz, \kt, z)
                       \cdot \widetilde{\uv}^*_{\alpha, \text{C}} (\kzi{n}, \kt, z) \, , \notag
\end{eqnarray}
which obeys the commutation relations (\ref{comm1}).

In a similar fashion, we derive the electric field valid outside the cavity region only.
For this, we first express $\ev^\pf_\text{O}(\rv,t)$ [see Eq.~(\ref{elf})] in the limit
$\eta \rightarrow \, \infty$ [see Eq.~(\ref{rt-limit})],
\begin{eqnarray}\label{elf-ol}
\ev^\pf_\text{OL}(\rv,t) &=& \sum_\alpha \intdh \kv \sqrt{\frac{2 \hbar \; \omega_\kv}{\epsilon (2 \pi)^3}}
                            \times \notag \\
                        && \qquad \widetilde{\uv}_{\alpha, \text{O}}(\kv, z) \, a_\alpha(\kv)  \,
                           e^{\im (\kt \cdot \rv - \omega_\kv t)} \, ,
\end{eqnarray}
where the reduced mode functions
\begin{eqnarray}
\widetilde{\uv}_{\alpha, \text{O}} (\kv, z) \equiv
\lim_{\eta \rightarrow \, \infty} \uv_{\alpha, \text{O}}(\kv, z) =
                      \widetilde{\uv}_{\alpha, \text{C}} (\kv, z - \ell) \notag
\end{eqnarray}
form an orthonormal set
\begin{equation}\label{orth-set2}
\int dz \, \widetilde{\uv}_{\alpha, \text{O}}
(\kz, \kt, z) \cdot \widetilde{\uv}^*_{\alpha^\prime, \text{O}}
(\kz^\prime, \kt, z) = 2 \pi \; \delta_{\alpha, \alpha^\prime} \, \delta (\kz - \kz^\prime) \, ,
\end{equation}
with the integration being restricted to $0 < z < \infty$.

By following the same approach as before,
we keep Eq.~(\ref{elf-ol}) with replacing $a_\alpha(\kv)$ by operator $b_\alpha(\kv)$
\begin{eqnarray}\label{elf-op}
\ev^\pf_\text{OP}(\rv,t) &=& \sum_\alpha \intdh \kv \sqrt{\frac{2 \hbar \; \omega_\kv}{\epsilon (2 \pi)^3}}
                            \times \notag \\
                        && \qquad \widetilde{\uv}_{\alpha, \text{O}}(\kv, z) \, b_\alpha(\kv)  \,
                           e^{\im (\kt \cdot \rv - \omega_\kv t)} ,
\end{eqnarray}
where we interpret $b_\alpha(\kv)$ as the photon annihilation operator of
the (reservoir) modes outside the cavity. This operator satisfies the usual
commutation relations
\begin{subequations}\label{comm2}
\begin{eqnarray}
&& [ b_\alpha(\kv), b^\dagger_{\alpha^\prime}(\kv^\prime) ]
        = \delta_{\alpha, \alpha^\prime} \, \delta \left( \kv - \kv^\prime \right), \label{comm2a} \\
&& \hspace{1cm} [ b_\alpha(\kv), b_{\alpha^\prime}(\kv^\prime) ] = 0 \, . \label{comm2b}
\end{eqnarray}
\end{subequations}

In order to find the reservoir photon operator $b_\alpha(\kv)$ in terms of
$a_\alpha(\kv)$, we multiply (scalarly) both Eqs.~(\ref{elf}) and (\ref{elf-op})
by $\widetilde{\uv}^*_{\alpha, \text{O}} (\kz^\prime, \kt, z)$ and integrate
them over $z$ from $0$ to $\infty$ using the property (\ref{orth-set2}). We equate
the resulting expressions and solve them for the operator $b_\alpha(\kv)$,
\begin{eqnarray}\label{b-op}
&& b_\alpha(\kv) = \frac{1}{2 \, \pi} \intdh \kz^\prime \sqrt{\frac{\omega^\prime_\kv}{\omega_\kv}}
                   \, a_\alpha (\kz^\prime, \kt) \times \\
                 && \qquad e^{\im ( \omega_\kv - \omega^\prime_\kv ) \, t}
                    \intd z \, \uv_{\alpha, \text{O}} (\kz^\prime, \kt, z)
                    \cdot \widetilde{\uv}^*_{\alpha, \text{O}} (\kz, \kt, z) \, , \notag
\end{eqnarray}
where $\omega^\prime_\kv \equiv c \sqrt{ \kz^\prime + k^{\, 2} }$ and which obeys the
commutation relations (\ref{comm2}).

The relations (\ref{orth-set1}) and (\ref{orth-set2}) along with the
commutation relations (\ref{comm1}) and (\ref{comm2}) imply that the
electric fields $\ev_\text{CP}(\rv,t)$ and $\ev_\text{OP}(\rv,t)$,
along with the respective magnetic fields, enable to describe any
physically achievable configuration of the electromagnetic field
inside and outside the planar resonator, respectively. Except for
the region filled by the semitransparent mirror, therefore, the
operators $c_{\alpha,n}(\kt)$ and $b_\alpha(\kv)$ cover the entire
continuum of Fock spaces spanned by the global operator $a_\alpha
(\kv)$. In other words, this operator can be expanded as
\begin{equation}\label{expans0}
a_\alpha (\kv) = \sum_n A_{\alpha, n}(\kv) \, c_{\alpha, n}(\kt)
                    + \intdh \kz^\prime B_\alpha(\kv, \kz^\prime) b_\alpha (\kz^\prime, \kt),
\end{equation}
where $A_{\alpha, n}(\kv)$ and $B_\alpha(\kv, \kz^\prime)$ are defined
by the means of commutators
\begin{subequations}\label{proj0}
\begin{eqnarray}
&& A_{\alpha, n}(\kv) = \left[ a_\alpha (\kv), \int d \kt^\prime \, c^\dag_{\alpha,n}(\kt^\prime) \right] \, , \\
&& B_\alpha(\kv, \kz^\prime) = \left[ a_\alpha (\kv), \int d \kt^\prime \, b^\dag_\alpha(\kv^\prime) \right] \, .
\end{eqnarray}
\end{subequations}

Since the global photon operator obeys the eigenoperator equation
$\left[ a_\alpha (\kv), H_\text{F} \right] = \hbar \, \omega_{\kv} \, a_\alpha (\kv)$
[see Eq.~(\ref{ham-t})], the expansion (\ref{expans0}) can be traced back to
Fano's diagonalization technique utilized in Ref.~\cite{pr124} to
analyze a coupled bound-continuum system. In the framework of cavity QED,
this technique has been exhaustively studied in Ref.~\cite{BR},
where $a_\alpha (\kv)$, $c_{\alpha, n}(\kt)$, and $b_\alpha(\kv)$
were identified as the dressed, bare (or quasi-cavity), and reservoir
photon operators, respectively (see also Ref.~\cite{oc68}).

\subsection{The case of a high but finite reflectivity}

We found above that $a_\alpha (\kv)$ can be expressed with the help of operators
$c_{\alpha,n}(\kt)$, $b_\alpha(\kv)$ and functions (\ref{proj0}). This expected result,
obtained for a lossless resonator, is based on the ability to describe any attainable
electromagnetic field configuration inside or outside the resonator using the expression
(\ref{elf-cp}) or (\ref{elf-op}), respectively.

In this section, we show that the expansion (\ref{expans0}) holds true also in the case
of a high but finite reflectivity, i.e., a leaky cavity. In order to proceed, we replace
the functions $R_\alpha(\kv)$ and $T_\alpha(\kv)$ in (\ref{elf-cp}) and (\ref{elf-op})
by the expressions (\ref{new}). The
structure of $\ev^\pf_\text{CP}(\rv,t)$ and $\ev^\pf_\text{OP}(\rv,t)$ implies that the
reduced mode functions $\widetilde{\uv}_{\alpha, \text{C}}(\kzi{n}, \kt,z)$,
$\widetilde{\uv}_{\alpha, \text{O}}(\kv,z)$ and the orthogonality relations (\ref{orth-set1}),
(\ref{orth-set2}) remain unchanged. The operators $c_{\alpha,n}(\kt)$ and
$b_\alpha(\kv)$, in contrast, include implicitly the reflectivity and transmissivity
by means of the spectral response function $L_\alpha(\kv)$ [see Eqs.~(\ref{lp-def})]. In
the case of a leaky cavity with $\tau \ll 1$, to a good approximation, this response
function takes the form
\begin{eqnarray}\label{l-def+}
L(\omega, k) \cong \frac{c}{2 \, \ell}
   \sum_{n=0}^\infty \frac{- \tau}{\omega - \omega_{n, k} + \im c \, \tau^2 / (4 \, \ell)} \, ,
\end{eqnarray}
where, without loss of generality, we replaced $\kz$ by the (frequency valued) parameter 
$\omega$ divided  by $c$. In the denominator of this expression, moreover, we have imposed 
the $k$ dependence by means of the quasi-mode frequency $\omega_{n, k}$. In the limit 
of vanishing $k$, the resulting function reduces to Eq.~(9.49) derived in Ref.~\cite{D} 
for the case of an one-dimensional leaky cavity, where the contribution of continuous 
and unconfined modes has been omitted.

By inserting (\ref{l-def+}) in Eqs.~(\ref{c-op}) and (\ref{b-op}) with replacement
$\kz \rightarrow \omega / c$, we compute explicitly $c_{\alpha,n}(\kt)$ and $b_\alpha(\kv)$,
\begin{eqnarray}
&& c_{\alpha, n}(\kt) = \int^\prime d \omega \, \frac{- \tau /(2 \, \sqrt{\pi \, \ell})}
                       {\omega - \omega_{n, k} + \im c \, \tau^2 /(4 \, \ell)} \,
                        a_{\alpha}(\omega, \kt) \, , \label{c-op1} \\
\label{b-op1}
&& b_{\alpha}(\kv) = \int^\prime d \omega \, \, a_{\alpha}(\omega, \kt)
                        \left[ \delta \left( c \, \kz - \omega \right) \right. \\
                && \quad \quad + \left. \lim_{\xi \rightarrow 0^+} \frac{1}{c \, \kz - \omega - \im \xi} \,
                    \sum_{n=0}^\infty \frac{- \tau^2 / (4 \, \pi \, \ell)}
                               {\omega - \omega_{n, k} + \im c \, \tau^2 / (4 \, \ell)} \right], \notag
\end{eqnarray}
where the global photon operator fulfills the relations
\begin{eqnarray}
&& [ a_\alpha(\omega, \kt), a^\dagger_{\alpha^\prime}(\omega^\prime, \kt^\prime) ]
        = c \, \, \delta_{\alpha, \alpha^\prime} \, \delta \left( \omega - \omega^\prime \right) \,
          \delta \left( \kt - \kt^\prime \right), \notag \\
&& \hspace{2cm} [ a_\alpha(\omega, \kt), a_{\alpha^\prime}(\omega^\prime, \kt^\prime) ] = 0 \, . \notag
\end{eqnarray}
\begin{figure}
\begin{center}
\includegraphics[width=0.45\textwidth]{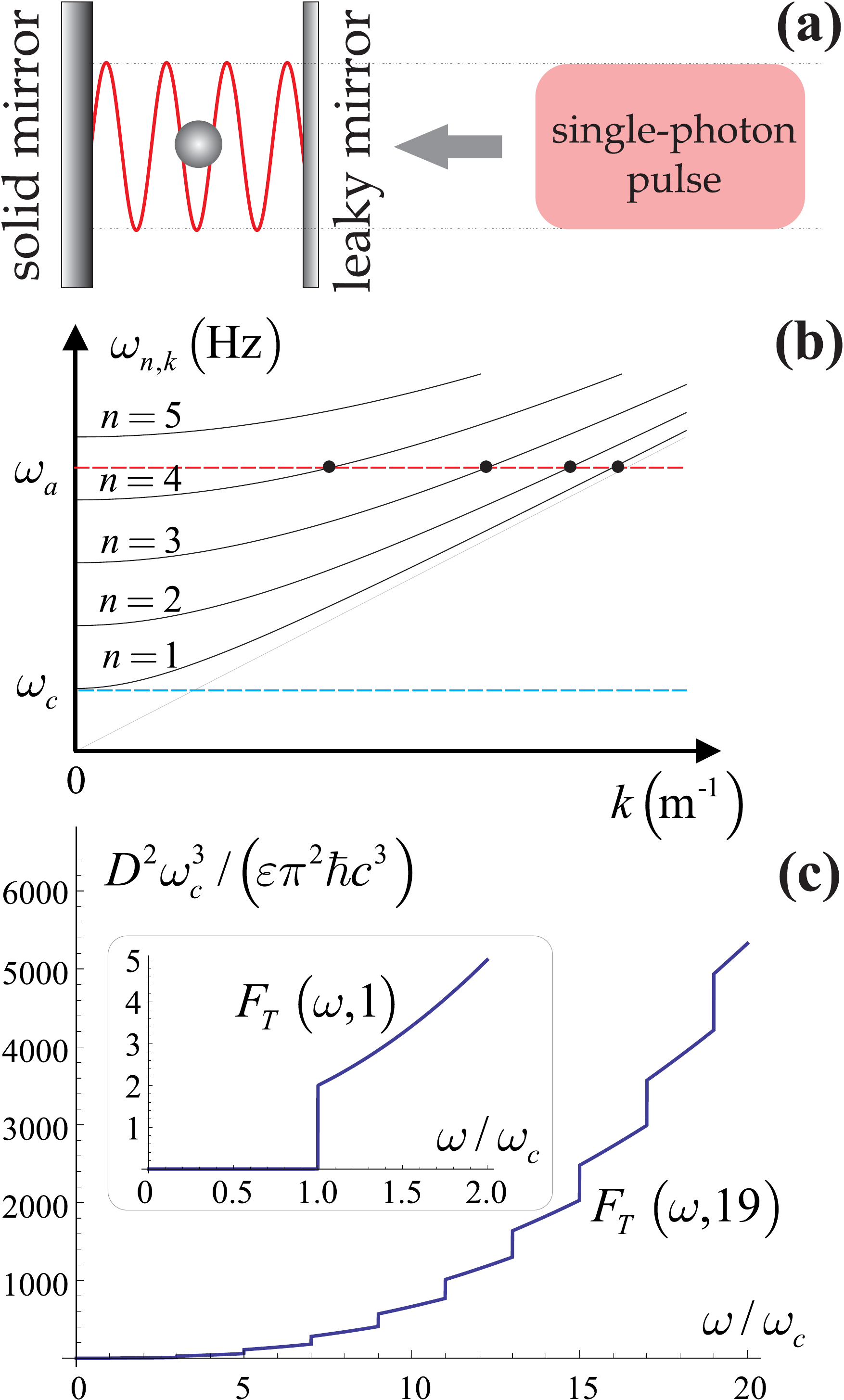} \\
\caption{(Color online) (a) Schematic view of the experimental setup that realizes
the proposed scenario. (b) Series of branches which characterize the dispersion relation
$\omega_{n, k} = c \sqrt{ \kzi{n}^2 + k^{2} }$ associated with the frequency of cavity
quasi-modes. (c) The total form-factor (\ref{fft}) for $N=1$ and $N=19$ being summed over
the polarizations. See text for details.}
\label{fig2}
\end{center}
\end{figure}

We show below that these operators fulfill the commutation relations 
(\ref{comm1}) and (\ref{comm2}), respectively. Using the same arguments 
as in the previous section, i.e., the possibility to describe any 
configuration of the electromagnetic field using (\ref{elf-cp}) and 
(\ref{elf-op}), we conclude that the expansion
\begin{equation}\label{expans}
a_\alpha (\omega, \kt) = \sum_n A_n(\omega, k) \, c_{\alpha,n}(\kt)  + \intdh \kz \, B(\omega, \kz) \, b_\alpha (\kv)
\end{equation}
replaces Eq.~(\ref{expans0}) in the case of a leaky cavity, where
\begin{subequations}\label{proj}
\begin{eqnarray}
&& A_n(\omega, k) = \frac{- \tau /(2 \, \sqrt{\pi \, \ell})}
                      {\omega - \omega_{n, k} - \im c \, \tau^2 /(4 \, \ell)} \, , \label{proj1} \\
&& B(\omega, \kz) = \delta \left( c \, \kz - \omega \right) \notag \\
                  && \hspace{1cm} + \lim_{\xi \rightarrow 0^+} \frac{\tau / (2 \, \sqrt{\pi \, \ell})}
                   {c \, \kz - \omega + \im \xi} \, \sum_{n=0}^\infty A_n(\omega, k) \, . \qquad \label{proj2}
\end{eqnarray}
\end{subequations}

In order to show that $c_{\alpha, n}(\kt)$ and $b_{\alpha}(\kv)$ satisfy the
commutation relations, we first use Eqs.~(\ref{c-op1}) and (\ref{proj1}), for which
the commutator (\ref{comm1a}) reduces to the expression
\begin{equation}\label{comm1+}
\delta_{\alpha, \alpha^\prime} \, \delta_{n, n^\prime} \,
\delta \left( \kt - \kt^\prime \right) \int^\prime d \omega \, c \, |A_n(\omega, k)|^2 .
\end{equation}
Now we use Eqs.~(\ref{b-op1}) and (\ref{proj2}), for which the commutator (\ref{comm2a})
reduces to the expression
\begin{equation}\label{comm2+}
\delta_{\alpha, \alpha^\prime} \, \delta \left( \kt - \kt^\prime \right)
\int^\prime d \omega \, c \, B(\omega, \kz^\prime) \, B^*(\omega, \kz) \, .
\end{equation}
It can be readily checked that the integral in (\ref{comm1+}) is equal to one,
while the integral in (\ref{comm2+}) is equal to $\delta \left( \kz - \kz^\prime \right)$
up to the contribution $\mathcal{O}(\tau^4)$, which is negligibly small due to the
(leaky cavity) condition $\tau \ll 1$.

\section{Form-factors and the cavity-QED parameters}

In the introduction, we explained that an interacting system of three constituents:
(i) an atom coupled to (ii) a planar resonator, and (iii) an input pulse that penetrates
the resonator from outside, can exhibit Rabi oscillations reproducing the
cavity-QED evolution. In Fig.~\ref{fig2}(a) we display the experimental setup that could
realize this scenario. In this setup, an atom at rest is located inside the planar
resonator, while the single-photon wave packet
\begin{equation}\label{pulse0}
\ket{\text{in}} = \sum_\alpha
           \int d \rv \, \psi(z) \, \varphi_\alpha(x,y) \, \tilde{b}^\dag_\alpha (\rv) \, \ket{\text{vac}}
\end{equation}
characterized by a non-trivial spatial distribution penetrates the resonator at
normal incidence. Here $\tilde{b}_\alpha (\rv)$ denotes the Fourier transform of
$b_\alpha (\kv)$, while $\ket{\text{vac}}$ is the photon field vacuum state, such
that $a_\alpha (\kv) \ket{\text{vac}} = 0$. The functions $\psi(z)$ and
$\varphi_\alpha(x,y)$ describing the spatial distribution of the light pulse are
normalized, such that
\begin{equation}\label{norm0}
\int dz \, |\psi(z)|^2 = \sum_\alpha \int dx \, dy  \, |\varphi_\alpha(x,y)|^2 = 1 \, ,
\end{equation}
and where $|\psi(z)|^2$ is localized in the $z>0$ region.

Similar to the standard cavity-QED, the coupled atom-cavity-pulse system is described by
a set of parameters which characterize completely the coherent (unitary) and incoherent
(non-unitary) parts
of its evolution. Before we identify this set, we formulate two criteria by which we determine
these parameters. First, this set should contain only three elements with the same
physical meaning as the cavity-QED parameters $(g, \, \kappa, \, \gamma)$. This requirement
would establish a one-to-one correspondence to the conventional cavity-QED framework.
Secondly, these parameters have to depend not only on the atom and cavity characteristics
(as in cavity-QED) but also on the lateral profile $\varphi_\alpha(x,y)$ of the single-photon
pulse (\ref{pulse0}).

The identification of these parameters, framed by the above two
criteria, was proposed by Koshino in Ref.~\cite{K}. In this
reference, the author observed that the atom-field coupling
extracted from the total Hamiltonian including an atom, the light
field, and the atom-cavity interaction encodes the entire triplet of
parameters. Using the so-called form-factor formalism that gives a
proper framework to isolate and study various couplings of a given
Hamiltonian, Koshino calculated the cavity-QED parameters in
question and demonstrated that the atomic decay rate $\gamma$
becomes considerably suppressed once the lateral profile
$\varphi_\alpha(x,y)$ of the input single-photon pulse is
appropriately tailored.

However, Koshino introduced four simplifying assumptions in his
framework, namely, (i) the evolution of the coupled
atom-cavity-pulse system was described by an \textit{ad hoc}
Hamiltonian, (ii) the light field had only one (fixed) polarization,
(iii) the atom was described by an averaged (in space) dipole, while
(iv) the planar resonator accommodated only one atomic wavelength.
In the generalized framework we derived in the previous section, the
first two simplifications have been already excluded. In this
section, we avoid the remaining two assumptions and generalize, in
this way, the paper of Koshino.

\subsection{Input light-pulse coupled to the resonator}

The light pulse (\ref{pulse0}) can be expressed in $\kv$-space as
\begin{equation}\label{pulse1}
\ket{\text{in}} = \sum_\alpha \intd \kv \, \widetilde{\psi}(\kz) \, \widetilde{\varphi}_\alpha(\kt) \,
                  b^\dag_\alpha(\kv) \ket{\text{vac}} \, ,
\end{equation}
where $\widetilde{\psi}(\kz)$ and $\widetilde{\varphi}_\alpha(\kt)$ are the
Fourier transforms of $\psi(z)$ and $\varphi_\alpha(x,y)$, respectively, and
describe the frequency distribution of the input light pulse in $\kv$-space.
Similar to Eq.~(\ref{norm0}), these two functions are normalized
\begin{equation}\label{norm1}
\intd \kz \, |\widetilde{\psi}(\kz)|^2 =
                                \sum_\alpha \int d \kt \, |\widetilde{\varphi}_\alpha(\kt)|^2 = 1 \, .
\end{equation}
In the conventional approach \cite{BR, pra42}, a single-photon state with
a non-trivial frequency distribution is typically given by the expression
$\ket{1} = \int d \omega_\kv \, \mathcal{G}(\omega_\kv) \, b^\dag(\omega_\kv) \ket{\text{vac}}$,
where $\omega_\kv$ is proportional to the modulus of $\kv$. In our case,
however, the frequency distribution is given by $\widetilde{\psi}(\kz)$ and
$\widetilde{\varphi}_\alpha(\kt)$ depending on different wave-vector components,
while the integration is performed over the $\kv$-space.

Using (\ref{b-op1}) and (\ref{proj}), we express (\ref{pulse1}) in the form
\begin{equation}\label{pulse2}
\ket{\text{in}} = \frac{1}{c} \int d \omega \, \widetilde{\psi}(\omega) \, 
                  d^\dag (\omega, \widetilde{\varphi}) \ket{\text{vac}} \, ,
\end{equation}
that is similar to the conventional single-photon state $\ket{1}$ shown above,
and where we introduced the \textit{pulse} operator
\begin{eqnarray}\label{d-op}
d (\omega, \widetilde{\varphi}) &\equiv& \sum_\alpha \int d \kt \, \widetilde{\varphi}^{\, *}_\alpha(\kt) \times \notag \\
              &&  \left[ 1 + \im \tau \sqrt{\frac{\pi}{\ell}} \sum_{n=0}^\infty  A^*_n(\omega, k) \right]
                  a_\alpha(\omega, \kt) \, , \qquad
\end{eqnarray}
that depends on the lateral profile $\widetilde{\varphi}_\alpha$. This operator,
moreover, fulfills the commutation relations
\begin{subequations}\label{comm3}
\begin{eqnarray}
&& [ d(\omega, \widetilde{\varphi}), d^\dagger(\omega^\prime, \widetilde{\varphi}) ]
        = c \, \, \delta \left( \omega - \omega^\prime \right); \\
&& \hspace{1cm} [ d(\omega, \widetilde{\varphi}), d(\omega^\prime, \widetilde{\varphi}) ] = 0 \, ,
\end{eqnarray}
\end{subequations}
and exhibits the properties
\begin{subequations}\label{props}
\begin{eqnarray}
d^\dagger(\omega, \widetilde{\varphi}) \ket{\text{vac}}
                     &=& \widetilde{\psi}^*(\omega) \, \ket{\text{in}} \, , \label{prop1} \\
d(\omega, \widetilde{\varphi}) \ket{\text{in}}
                     &=& \widetilde{\psi}(\omega) \, \ket{\text{vac}} \, . \label{prop2}
\end{eqnarray}
\end{subequations}

These properties suggest
that $d^\dag(\omega, \widetilde{\varphi})$ creates a single-photon state (\ref{pulse1})
weighted by $\widetilde{\psi}^*(\kz)$, while $d(\omega, \widetilde{\varphi})$
annihilates the respective state resulting into the vacuum state weighted by
$\widetilde{\psi}(\kz)$. In order to complete the derivations in this subsection,
we invert the relation (\ref{d-op})
\begin{equation}\label{invert}
a_\alpha(\omega, \kt) = \widetilde{\varphi}_\alpha(\kt)
                \left[ 1 - \im \tau \sqrt{\frac{\pi}{\ell}} \sum_{n=0}^\infty  A_n(\omega, k) \right]
                d(\omega, \widetilde{\varphi}) \, ,
\end{equation}
where we used Eq.~(\ref{norm1}) along with the relation
\begin{equation}\label{ident}
\im \tau \sqrt{\frac{\pi}{\ell}} \, \left| A_n(\omega, k) \right|^2 = A^*_n(\omega, k) - A_n(\omega, k) \, .
\end{equation}
\subsection{Atom coupled to the resonator}

In this section, we derive the Hamiltonian that governs the evolution of the intra-cavity
field coupled to a two-level atom by using the electric field (\ref{elf-cp}) and the
photon field operator (\ref{c-op1}) of a leaky cavity. We consider an atom at rest
inside the resonator as displayed in Fig.~\ref{fig2}(a). The internal structure of the
atom is completely characterized by the states $\ket{g}$ (ground) and $\ket{e}$ (excited),
which fulfill the usual orthogonality and completeness relations. We recall that the
cavity quasi-modes are characterized by the frequency
$\omega_{n, k}$ that has both discrete and continuous
contributions. These quasi-modes are grouped by branches indexed by $n$ as can be seen
in Fig.~\ref{fig2}(b), where $\omega_c \equiv \omega_{1,0} = c \, \kzi{1}$ defines the
lower \textit{cut-off} frequency. We assume that the atomic transition frequency 
$\omega_a$ is equal or above this lower cut-off frequency, such that the atom couples 
at least to one quasi-mode of the resonator.

We choose the position of atomic center-of-mass $\rv^\prime = \{ 0, 0,-\ell/2 \}$ and switch
to the Schr{\"o}dinger picture, in which the electric field (\ref{elf-cp}) is time-independent.
In the dipole approximation, the Hamiltonian ($n=1, \ldots, N$)
\begin{eqnarray}\label{ac-ham1}
H_\text{AC} &=& - \mathcal{Q} \, \mathbf{r} \cdot
              \left[ \ev^\pf_\text{CP}(\rv^\prime) + \ev^\nf_\text{CP}(\rv^\prime) \right] \notag \\
            &=& - \mathcal{D} \left( \sigma^\dag + \sigma \right)
               \sum_{\alpha,n}^{N} \int d \kt \, \sqrt{\frac{\hbar \; \omega_{n,k}}{\epsilon (2 \pi)^2 \, \ell}}
               \times \notag \\
&& \left[ \vecP \cdot \widetilde{\uv}_{\alpha, \text{C}} \left(\kzi{n}, \kt, -\frac{\ell}{2} \right)
          c_{\alpha, n}(\kt) + H.c. \right] \quad
\end{eqnarray}
describes the electric-dipole coupling between a two-level atom and $N$ cavity quasi-modes,
where $N$ is the number of intersection points between $\omega_a$ and the branches of
$\omega_{n, k}$ [see Fig.~\ref{fig2}(b)]. In the above Hamiltonian,
$\mathcal{Q}$ is the electric charge, $\sigma = \ket{g} \bra{e}$ is the atomic
(excitation) lowering operator. We also introduced the notation
$\bra{g} \mathcal{Q} \, \mathbf{r} \ket{e} \equiv \mathcal{D} \, \vecP$ with $\mathcal{D}$
being the (real) dipole matrix element of the atomic transition and $\vecP$ being the unit
real vector that determines the polarization of transition.

In the rotating-wave approximation, we express the Hamiltonian (\ref{ac-ham1}) in the form
\begin{equation}\label{ac-ham2}
H_\text{AC} = \hbar \, \sum_{\alpha,n}^{N} \int d \kt
  \left[ \lambda_{\alpha, n}(\kt) \, \sigma^\dag \, c_{\alpha, n}(\kt) + H.c. \right] \, ,
\end{equation}
where we introduced the atom-field coupling
\begin{equation}\label{lambda}
\lambda_{\alpha, n}(\kt) \equiv - \mathcal{D} \sqrt{\frac{\omega_{n,k}}{\epsilon (2 \pi)^2 \, \ell \, \hbar}} \;
                           \vecP \cdot \widetilde{\uv}_{\alpha, \text{C}} \left( \kzi{n}, \kt, -\frac{\ell}{2} \right) \, .
\end{equation}
Using (\ref{ham-t}) and (\ref{ac-ham2}), we compose the total Hamiltonian
\begin{eqnarray}\label{tot-ham0}
H_\text{T} &=& H_\text{F} + H_\text{A} \notag \\
           &+& \hbar \, \sum_{\alpha,n}^{N} \int d \kt
             \left[ \lambda_{\alpha, n}(\kt) \, \sigma^\dag \, c_{\alpha, n}(\kt) + H.c. \right]
\end{eqnarray}
that governs the evolution of a coupled atom-field system, and where
$H_\text{A} = \hbar \, \omega_a \, \sigma_z /2$ denotes the atomic Hamiltonian.
Using (\ref{c-op1}) and (\ref{proj1}), furthermore, the above Hamiltonian takes
the (first) equivalent form
\begin{eqnarray}\label{tot-ham1}
H_\text{T} = H_\text{F} + H_\text{A} &+& \hbar \, \sum_{\alpha,n}^{N} \int^\prime d \omega \, d \kt
                                         \left[ \lambda_{\alpha, n}(\kt) \right. \times \notag \\
           && \left. A^*_n(\omega, k) \, \sigma^\dag \, a_{\alpha}(\omega, \kt) + H.c. \right]. \qquad
\end{eqnarray}
Using Eqs.~(\ref{invert}) and (\ref{ident}), finally, we express the above
Hamiltonian in the (second) equivalent form
\begin{eqnarray}\label{tot-ham2}
H_\text{T} = H_\text{F} + H_\text{A} &+& \hbar \, \sum_{\alpha,n}^{N} \int^\prime d \omega \, d \kt
             \left[ \lambda_{\alpha, n}(\kt) \, \widetilde{\varphi}_\alpha(\kt) \right. \times \notag \\
           && \left. A_n(\omega, k) \, \sigma^\dag \, d(\omega, \widetilde{\varphi}) + H.c. \right]. \qquad
\end{eqnarray}
\subsection{Form-factors and the cavity-QED parameters}

Following the approach of Koshino, we define the \textit{total} form-factor
\begin{eqnarray}\label{fft}
F_{T}(\omega, N) &\equiv& \frac{1}{c^3 \, \hbar^2 } \sum_\alpha \int d \kt \left| \bra{\text{vac}, e}
                      H_{\text{T}} \, a^\dag_{\alpha}(\omega, \kt) \ket{\text{vac}, g} \right|^2 \notag \\
                      &=& \frac{1}{c} \sum_\alpha \int d \kt \left| \sum_{n}^{N} \lambda_{\alpha, n}(\kt) \,
                      A^*_n(\omega, k) \right|^2 \qquad
\end{eqnarray}
that isolates the coupling between the atom and global photon field $a_{\alpha}(\omega, \kt)$.

Assuming that the atomic dipole $\vecP$ lies in the plane parallel to the mirrors,
i.e., $\vecP \cdot \hz = 0$, we calculate analytically the total form-factor (\ref{fft})
\begin{subequations}
\begin{eqnarray}
F_{T}(\omega, N) &=& \frac{\mathcal{D}^2 \, \omega_c^3}{\epsilon \, \pi^2 \hbar \, c^3}
  \sum_{n, \text{odd}}^N \left( n^2 + \frac{\omega^2}{\omega_c^2} \right) \times \notag \\
  && \left( \frac{1}{2} + \frac{1}{\pi} \, \arctan \left[ \frac{4 \, \pi}{\tau^2}
     \left( \frac{\omega}{\omega_c} - n \right) \right] \right) \label{fft-e} \qquad \\
  &\equiv& \frac{\mathcal{D}^2 \, \omega_c^3}{\epsilon \, \pi^2 \hbar \, c^3} \,
           F^\circ_{T} \left( \frac{\omega}{\omega_c}, N \right) \, , \label{fft-0}
\end{eqnarray}
\end{subequations}
that has the units of frequency, while the summation over
$n$ is only over the odd and positive integer values. We display in
Fig.~\ref{fig2}(c) the total form-factor $F_{T}(\omega, N)$, for $N
= 1$, $N = 19$ and $\tau = 10^{-3}$. It is clearly seen that the
total form-factor depends on the square of $\omega$ and has a
steplike behavior at the points $\omega = n \, \omega_c$ ($n =
1,3,5, \ldots$). This figure displays the spectral mode density of a
lossless (1D confined) planar resonator \cite{kav, D, MF} and,
therefore, it reveals the physical meaning of the total form-factor.

The chosen value of $\tau = 10^{-3}$ is compatible with the
assumptions of Sec.~II.A, by which a leaky mirror deviates only
slightly from a perfect one. Throughout this paper, therefore, we
consider this specific value to evaluate various expressions
involving $\tau$. We remark, moreover, that the total form-factor
derived by Koshino (see Eq.~(15) in Ref.~\cite{K}) is proportional
to $F_{T}(\omega, 1)$, however, with an excluded contribution of
$\omega^2 / \omega_c^2$ since the $\parallel$-component of light
polarization was disregarded.

Although we assumed that an input light-pulse penetrates the cavity as shown in
Fig.~\ref{fig2}(a), the total form-factor (\ref{fft-e}) is independent of the lateral
profile $\varphi_\alpha(x,y)$. We observe, however, that the total Hamiltonian
expressed in the form (\ref{tot-ham2}) contains two operator pairs
$\sigma \, d^\dag (\omega, \widetilde{\varphi})$ and
$\sigma^\dag d(\omega, \widetilde{\varphi})$, such that
\begin{subequations}
\begin{eqnarray}
&& \sigma \, d^\dag(\omega, \widetilde{\varphi}) \ket{\text{vac}, e}
   = \widetilde{\psi}^*(\omega) \, \ket{\text{in}, g} \, , \label{prop4} \\
&& \sigma^\dag d(\omega, \widetilde{\varphi}) \ket{\text{in}, g}
   = \widetilde{\psi}(\omega) \, \ket{\text{vac}, e} \, , \label{prop5}
\end{eqnarray}
\end{subequations}
These properties along with the atom-field coupling
$\lambda_{\alpha, n}(\kt) \, \widetilde{\varphi}_\alpha(\kt) \, A_n (\omega, k)$
suggest that, for an appropriately tailored $\widetilde{\varphi}_\alpha(\kt)$
[equivalently $\varphi_\alpha(x,y)$], the atom-field evolution governed by the
Hamiltonian (\ref{tot-ham2}) can resemble the evolution of a cavity-QED system,
that is
\begin{equation}\label{behav}
e^{-\frac{\im}{\hbar} H_\text{T} \, t} \ket{\text{in}, g} = \ket{\text{vac}, e} \, ,
\end{equation}
where the photon field (\ref{d-op}) plays the role of cavity photon field in
cavity-QED [see (\ref{jc})]. In other words, if the single-photon pulse (\ref{pulse0})
penetrates a planar resonator with an atom in the ground state, then (after a certain 
time interval $t$) this pulse can be completely absorbed by the atom.

Furthermore, we define the second form-factor
\begin{equation}\label{ffc0}
F_{C}(\omega, \widetilde{\varphi}, N)
        \equiv \frac{1}{c^3 \, \hbar^2} \left| \bra{\text{vac}, e}
        H_{\text{T}} \, d^\dag(\omega, \widetilde{\varphi}) \ket{\text{vac}, g} \right|^2 \, ,
\end{equation}
to which we refer below as the \textit{cavity} form-factor, since it contains the
coupling between an atom and the ($\widetilde{\varphi}_\alpha$-dependent) photon
field (\ref{d-op}) that might reproduce the cavity-QED evolution (\ref{behav}). By
inserting the total Hamiltonian (\ref{tot-ham2}) into the expression (\ref{ffc0}),
we readily obtain
\begin{equation}\label{ffc}
F_{C}(\omega, \widetilde{\varphi}, N) =
   \frac{1}{c} \left| \sum_{n, \alpha}^{N} \int d \kt \, \lambda_{\alpha, n}(\kt) \,
   \widetilde{\varphi}_\alpha(\kt) \, A_n \left( \omega, k \right) \right|^2 .
\end{equation}

The cavity-QED like behavior (\ref{behav}), exhibited by the atom-cavity-pulse system
with an appropriate input pulse, suggests that the cavity form-factor (\ref{ffc}) plays
the role of the spectral mode density corresponding to the completely (3D) confined
light. In a cavity-QED system with reasonable
small losses, in turn, this density produces a resonance peak centered around
$\omega_\circ$ and described by the Lorentzian
\begin{equation}\label{lorentzian}
\mathcal{L}(\omega) = \frac{\kappa}{2 \pi} \frac{g^2}{(\omega - \omega_\circ)^2 + \kappa^2 /4} \, ,
\end{equation}
where its area and the half-width are identified with $g^2$ and $\kappa$, respectively
\cite{kav, D, MF}. Using this analogy and provided that the cavity form-factor resembles
a sharply peaked resonance, we identify the atom-field coupling strength with the
expression
\begin{equation}\label{g-def}
g(\widetilde{\varphi}, N) \equiv
  \left( \int d \omega \, F_{C}(\omega, \widetilde{\varphi}, N) \right)^{\frac{1}{2}} \, ,
\end{equation}
while the cavity relaxation ratio $\kappa(\widetilde{\varphi}, N)$ is identified with the
half-width of $F_{C}(\omega, \widetilde{\varphi}, N)$.

We notice that in contrast to the total form-factor (\ref{fft}), the modulus in (\ref{ffc})
is moved outside the integral. With the help of Cauchy–-Schwarz inequality, this feature
leads to the relation
\begin{equation}\label{ineq}
F_{C}(\omega, \widetilde{\varphi}, N) \leq F_{T}(\omega, N) \, .
\end{equation}
Since the resonance peak (\ref{lorentzian}) describing the spectral mode density of a
cavity-QED system is below the (quadratically growing) curve given by
Eq.~(\ref{fft-e}) and corresponding to the spectral mode density of a planar resonator,
both above form-factors are in perfect agreement with the relation (\ref{ineq}) and the
present discussion.

The above relation suggests the third form-factor,
\begin{equation}\label{ffn}
F_{N}(\omega, \widetilde{\varphi}, N) \equiv
      F_{T}(\omega, N) - F_{C}(\omega, \widetilde{\varphi}, N) \, ,
\end{equation}
to which we refer below as the \textit{non-cavity} form-factor, since it gives the
difference between the spectral mode densities of (i) the (1D confined) atom-cavity
system and (ii) the atom-cavity-pulse system that behaves as a (3D confined)
cavity-QED system with losses. Since the cavity form-factor (\ref{ffc}) satisfies
the relation (\ref{ineq}) and encodes $g(\widetilde{\varphi}, N)$ and
$\kappa(\widetilde{\varphi}, N)$, we identify the expression (\ref{ffn}) with the
atomic decay rate,
\begin{equation}\label{gamma-def}
\gamma (\omega, \widetilde{\varphi}, N) \equiv
         F_{N}(\omega, \widetilde{\varphi}, N) \, .
\end{equation}

In accordance with the two criteria we formulated in the beginning of this section, we
defined three form-factors characterizing the coherent and incoherent parts of evolution
of the coupled atom-cavity-pulse system. Being defined in a similar fashion as in the
conventional cavity-QED, these form-factors enable us to compare
the overall performance of our setup to an arbitrary cavity-QED system. We recall that
the input state (\ref{pulse2}) is a single-photon state with a non-trivial frequency
distribution $\widetilde{\psi}(\omega)$, where the parameter $\omega$ contributes to all
the form-factors, while the lateral profile $\widetilde{\varphi}_\alpha(\kt)$ contributes
only to the cavity and non-cavity form-factors. In order to enhance the atom-field
interaction, in the next section, we determine the optimal frequency distribution 
$\widetilde{\psi}^\text{opt}$ and the optimal lateral profile 
$\widetilde{\varphi}^\text{opt}_\alpha$, for which the atomic decay rate vanishes.

\section{Analysis of lateral profiles}

\begin{figure}
\begin{center}
\includegraphics[width=0.475\textwidth]{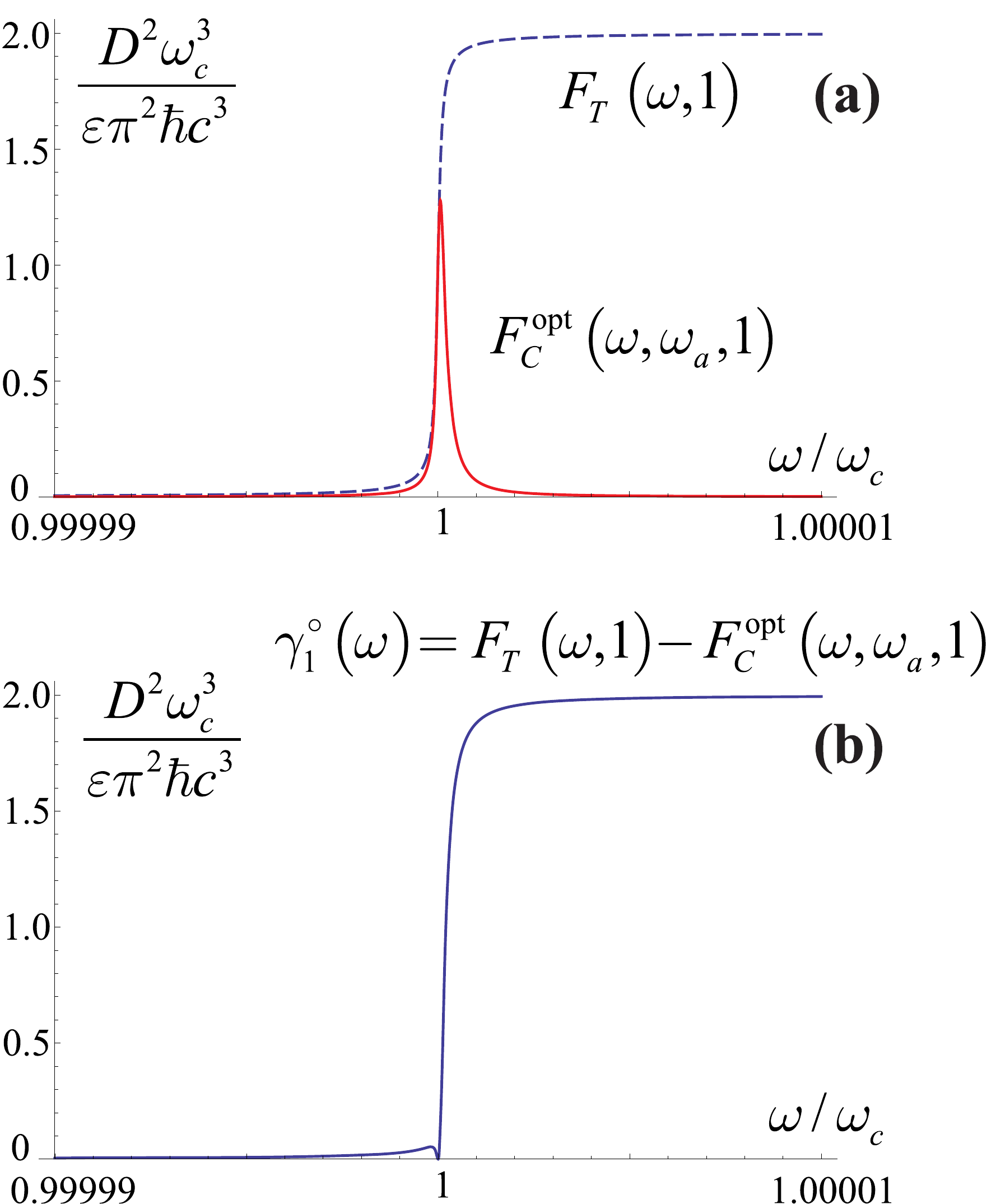} \\
\caption{(Color online) (a) The cavity (solid curve) and the total (dashed curve)
form-factors for $N=1$ and $f = \omega_a$. The cavity form-factor resembles a nice
Lorentzian bounded by the total form-factor. (b) The non-cavity form-factor for $N=1$
and $f = \omega_a$. See text for details.}
\label{fig3}
\end{center}
\end{figure}

In the previous sections, we identified the cavity-QED parameters which characterize
both coherent and incoherent parts of the evolution of a coupled atom-cavity-pulse
system with losses. We also suggested that the setup displayed in Fig.~(\ref{fig2})(a)
can reproduce the cavity-QED evolution once an appropriately tailored single-photon
pulse and a proper frequency distribution are provided at the input. In this section,
we demonstrate that a coupled atom-cavity-pulse system behaves as a cavity-QED system
and we evaluate the cavity-QED parameters by considering several predefined pulses.

\subsection{Optimal spatial distribution}

Before we evaluate cavity-QED parameters for a predefined single-photon pulse,
we determine first $\widetilde{\varphi}^\text{opt}_\alpha$ and
$\widetilde{\psi}^\text{opt}$, for which the atomic decay rate vanishes.
According to the definition (\ref{gamma-def}), we require the vanishing of the
left part of (\ref{ffn}). This leads to the equation
\begin{equation}\label{equal}
F_{T}(\omega, N) = F_{C}(\omega, \widetilde{\varphi}^\text{opt}_\alpha, N) \, ,
\end{equation}
which is an extreme case of Eq.~(\ref{ineq}). Using Eq.~(\ref{norm1}) and the cavity
form-factor (\ref{ffc}), we solve the above equation for the lateral profile. The
obtained solution
\begin{equation}\label{opt}
\widetilde{\varphi}^\text{opt}_\alpha(\kt, f, N) = \sum_n^N \frac{\lambda^*_{\alpha, n}(\kt)
                                 \, A^*_n \left( f, k \right)} {\sqrt{c \, F_{T}(f, N)}} \, ,
\end{equation}
fulfills Eq.~(\ref{norm1}) and depends on the parameters $f$ and $N$.

We insert the above optimal profile back into Eq.~(\ref{ffc}) and obtain the
optimal cavity form-factor
\begin{eqnarray}\label{ffc-opt0}
F^\text{opt}_{C}(\omega, f, N) &=&
   \frac{1}{c} \left| \sum_{n, \alpha}^{N} \int d \kt \, \lambda_{\alpha, n}(\kt) \times \right. \notag \\
&& \qquad \left. \widetilde{\varphi}^\text{opt}_\alpha(\kt, f, N) \,
                          A_n \left( \omega, k \right) \right|^2 . \qquad
\end{eqnarray}
Considering, as before, that the atomic dipole $\vecP$ lies in the plane parallel
to the mirrors, we calculate analytically the optimal cavity form-factor
(\ref{ffc-opt0}), which takes the form
\begin{equation}\label{ffc-opt}
F^\text{opt}_{C}(\omega, f, N) =
   \frac{\mathcal{D}^2 \, \omega_c^3}{\epsilon \, \pi^2 \hbar \, c^3}
   \left( \frac{\tau}{2 \, \pi} \right)^4
   \frac{\left| F^\circ_{C}(\frac{\omega}{\omega_c}, \frac{f}{\omega_c}, N) \right|^2}
        {F^\circ_{T} \left( \frac{f}{\omega_c}, N \right)} \, ,
\end{equation}
where $F^\circ_{T}(u, N)$ has been defined in (\ref{fft-0}), while
\begin{eqnarray}\label{ffc-opt1}
&& F^\circ_{C}(u, v, N) = \notag \\
&&   \sum_{n,\text{odd}}^N \left( n^2 \, \frac{\im 2 \, \pi^2 + 4 \, \pi \,
   \arctanh \left[ \frac{4 \pi \left( n - \frac{u + v}{2} \right)}
  {2 \pi (v - u) + \im \tau^2} \right]}{2 \, \pi (v - u) + \im \tau^2} \right. + \notag \\
&& \quad \left. \frac{\im 2 \, \pi^2 v^2 + \pi (u + v) \,
   \arctanh \left[ \frac{4 \pi \left( n - \frac{u + v}{2} \right)}
  {2 \pi (v - u) + \im \tau^2} \right]}{2 \, \pi (v - u) + \im \tau^2} \right) . \qquad
\end{eqnarray}

In Fig.~\ref{fig3}(a), we display $F^\text{opt}_{C}(\omega, \omega_a,
1)$ by a solid curve that corresponds to the situation, in which the
resonator accommodates just one wavelength associated with the atomic
transition frequency, that is $\ell = c \, \pi / \omega_a$. This
implies that only one quasi-mode couples to the atom, that is
$\omega_a = \omega_c$ and $N=1$. It is clearly seen that the solid
curve resembles a nice Lorentzian, while the peak of this Lorentzian
is bounded by the total form-factor $F_{T}(\omega, 1)$ (dashed curve)
in agreement with the relation (\ref{ineq}). This figure confirms the
identification of the cavity form-factor (\ref{ffc}) with the spectral
mode density of a cavity-QED system with losses. Moreover, this figure
reveals the role of parameter $f$ in Eq.~(\ref{ffc-opt}), and namely,
this parameter sets the central frequency of the resulting Lorentzian
(solid curve). In Fig.~\ref{fig3}(b), furthermore, we display the non-cavity
form-factor $F_{T}(\omega, 1) - F^\text{opt}_{C}(\omega, \omega_a, 1)$
that is identified with the (optimal) atomic decay rate $\gamma^\circ_1(\omega)$.
It can be clearly seen that the atomic decay is efficiently suppressed
in the region $\omega \leq \omega_c$. This restriction, in turn, suggests
the profile of the frequency distribution
$\widetilde{\psi}^\text{opt}(\omega)$.

We conclude that our system in Fig.~\ref{fig2}(a) behaves as a
cavity-QED system once the optimal single-photon pulse
\begin{equation}
\ket{\text{opt}_1} = \sum_\alpha \intd \kv \, \widetilde{\psi}^\text{opt}_{\omega_\circ}(\kz) \,
                                 \widetilde{\varphi}^\text{opt}_\alpha(\kt, \omega_a, 1) \,
                                 b^\dag_\alpha(\kv) \ket{\text{vac}} \, \notag
\end{equation}
is provided at the input, where $\widetilde{\psi}^\text{opt}_{\omega_\circ} (\omega)$
can be modeled by a narrow-band Gaussian distribution with the central
frequency $\omega_\circ$, such that $\omega_\circ \leq \omega_c$. Without
this input pulse, the spectral mode density describes an atom being weakly
coupled to the photon field confined in a planar resonator as seen
Fig.~\ref{fig2}(c) \cite{pra51, pra60}.

We assume now that the central frequency $\omega_\circ$ matches the atomic
transition frequency $\omega_a$($= \omega_c$) and we calculate $g^\circ_1$,
$\kappa^\circ_1$, and $\gamma^\circ_1(\omega_a)$ using the atomic data
\begin{equation}\label{data}
\lambda_a = 852 \; \text{nm} \, ; \quad \mathcal{D} = 4.48 \, \mathcal{Q} \, a_0 \, ,
\end{equation}
which correspond to the $D_2$-transition of a Cesium atom \cite{AD}, and where $a_0$ is
the Bohr radius. This atomic data, along with the Lorentzian (\ref{lorentzian}) plotted
in Fig.~\ref{fig3}(a), yields the cavity-QED parameters
\begin{equation}\label{params1}
\left( g^\circ_1, \kappa^\circ_1, \gamma^\circ_1(\omega_a) \right) =
        2 \pi \left( 49, 125, 0.2 \; 10^{-16} \right) \text{MHz} \, .
\end{equation}
We see that the cavity relaxation rate oversteps notably the atom-field coupling strength,
while the atomic decay rate is negligibly small if compared to both $g^\circ_1$ and
$\kappa^\circ_1$.

\begin{figure}
\begin{center}
\includegraphics[width=0.45\textwidth]{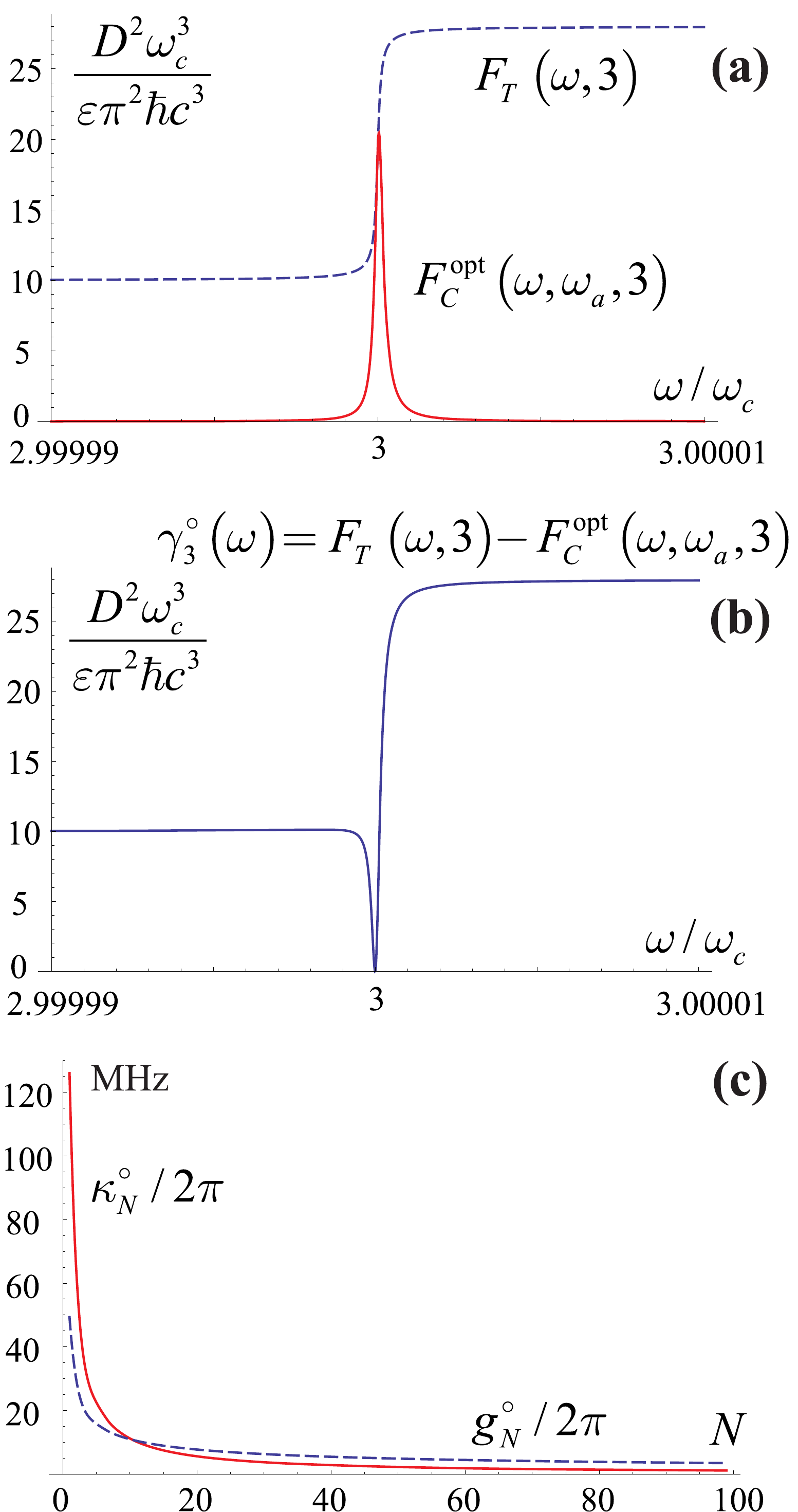} \\
\caption{(Color online) (a) The cavity (solid curve) and the total (dashed curve)
form-factors for $N=3$ and $f = 3 \, \omega_c$. The cavity form-factor resembles a nice
Lorentzian bounded by the total form-factor. (b) The non-cavity form-factor for $N=1$
and $f = 3 \, \omega_c$. (c) Cavity-QED parameters $g^\circ_N$ and $\kappa^\circ_N$ as
functions of $N$. See text for details.}
\label{fig4}
\end{center}
\end{figure}

The values (\ref{params1}) are obtained in the case when the
resonator accommodates just one atomic wavelength, such that the
atom is coupled to one single cavity quasi-mode, or equivalently,
$\omega_a = \omega_c$ and $N=1$. We stress that the suppression of
atomic decay for $\omega_\circ < \omega_c$ in a resonator accommodating
just one atomic wavelength was expected, since there are no
available quasi-modes below the cavity cut-off frequency to which an
input pulse can couple in order to facilitate the atomic emission.
We show below, however, that the inhibition of atomic emission occurs
in our setup even for $\omega_\circ > \omega_c$ in a resonator that
accommodates more than one atomic wavelength. This result cannot be
explained by the lack of available cavity quasi-modes and constitutes
a peculiar feature of the coupled atom-cavity-pulse system shown in
Fig.~\ref{fig2}(a).

In order to proceed, we consider the resonator that accommodates
three atomic wavelengths, such that the atom is coupled to three
quasi-modes, or equivalently, $\omega_a = 3 \, \omega_c$ and $N=3$.
In Figs.~\ref{fig4}(a) and (b), we display the cavity (solid curve)
and non-cavity form-factors, respectively. As in the previous case,
the cavity form-factor resembles a nice Lorentzian bounded by the
total form-factor (dashed curve), while the atomic decay rate is
suppressed inside a small window centered at $\omega = 3 \,
\omega_c$. Using the atomic data (\ref{data}) along with the
Lorentzian (\ref{lorentzian}) plotted in Fig.~\ref{fig4}(a), we
calculate $g^\circ_3$, $\kappa^\circ_3$, and
$\gamma^\circ_3(\omega_a)$, which take the values
\begin{equation}\label{params2}
\left( g_3^\circ, \kappa^\circ_3, \gamma^\circ_3(\omega_a) \right) =
        2 \pi \left( 21, 38, 0.3 \; 10^{-16} \right) \text{MHz} \, .
\end{equation}
If we compare these values to (\ref{params1}), we conclude that the cavity
relaxation rate oversteps slightly the coupling strength $g_3^\circ$, while
the atomic decay rate $\gamma^\circ_3(\omega_a)$ remains negligibly small.

To reveal the dependence of $g^\circ$ and $\kappa^\circ$ on $N$, we
consider the case when the resonator accommodates $N$ atomic
wavelengths, that is $\ell = N \, c \, \pi / \omega_a$. This implies
that the atom is coupled to $N$ cavity quasi-modes, such that
$\omega_a = N \, \omega_c$. With this in mind, the optimal pulse
\begin{equation}
\ket{\text{opt}_N} = \sum_\alpha \intd \kv \, \widetilde{\psi}^\text{opt}_{\omega_a}(\kz) \,
                                 \widetilde{\varphi}^\text{opt}_\alpha(\kt, \omega_a, N) \,
                                 b^\dag_\alpha(\kv) \ket{\text{vac}} \, \notag
\end{equation}
penetrates the resonator, where the central frequency $\omega_\circ$
matches the atomic transition frequency. As before, the cavity
form-factor yields a nice Lorentzian centered at $\omega = \omega_a$.
In Fig.~\ref{fig4}(c),
we display $g^\circ_N / 2 \, \pi$ (dashed curve) and $\kappa^\circ_N / 2 \,
\pi$ (solid curve) as functions of $N$, where the cavity length is
bounded by $N = 100$. This restriction corresponds to the length of
typical macroscopic resonators used in cavity-QED experiments in the
optical domain. It is clearly seen that in the region $N > 10$, the
cavity relaxation rate becomes slightly smaller than the respective
$g^\circ_N$ parameter leading, therefore, to the atom-field
evolution characterized by the inequality $g^\circ_N >
\kappa^\circ_N \gg \gamma^\circ_N(\omega_a)$. This regime ensures
that the energy exchange in the coupled atom-field system develops
faster than the losses due to the cavity relaxation and the atomic
decay. We remark that one reason, why the curves in
Fig.~\ref{fig4}(c) drop with growing $N$, is the fact that the
cut-off frequency that appears
in Eqs.~(\ref{fft}) and (\ref{ffc-opt}) drops with growing $\ell$,
which itself is proportional to $N$.

To summarize this section, we determined the optimal lateral profile of the
input pulse that ensures vanishing of the non-cavity form-factor identified with
the atomic decay rate. Using the cavity form-factor associated with the optimal
input pulse $\ket{\text{opt}_N}$, we studied the dependence of cavity-QED parameters
on the cavity
length that is proportional to the number of cavity quasi-modes coupled to the atom.
We confirmed that the atomic decay rate becomes dramatically suppressed once the
central frequency of pulse matches the atomic transition frequency. In
contrast to the atomic decay for $N=1$, which is suppressed for a rather large
window associated with frequency distribution $\widetilde{\psi}^\text{opt}(\omega)$,
the respective window for $N>1$ is much smaller and, therefore, hardly accessible
in practice.

\subsection{Hermite-Gaussian beam}

\begin{figure}
\begin{center}
\includegraphics[width=0.45\textwidth]{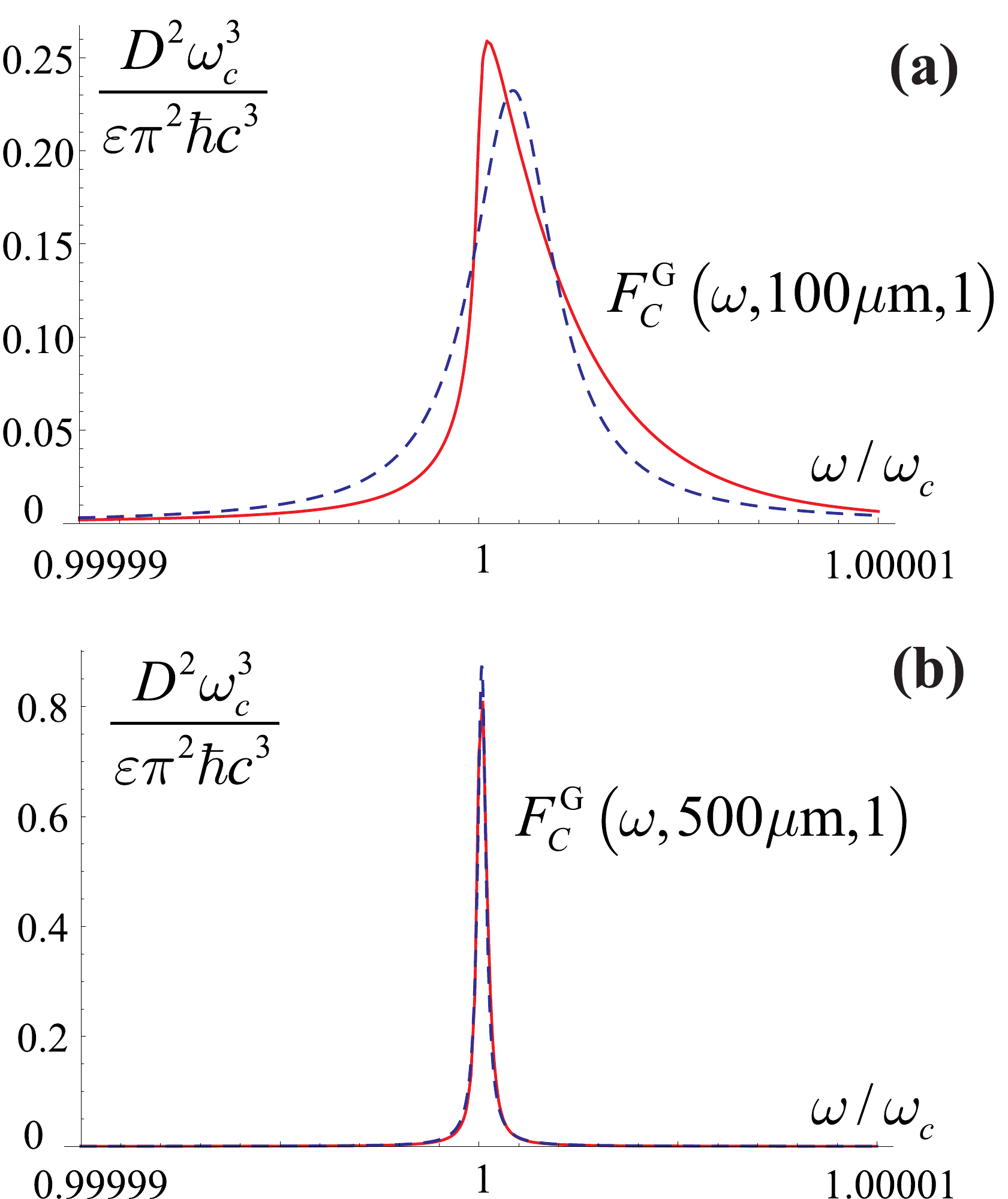} \\
\caption{(Color online) The cavity form-factor (solid curve) and the associated
Lorentzian (dashed curve) for $N=1$ and (a) $w = 100 \mu$m, (b) $w = 500 \mu$m.
See text for details.}
\label{fig5}
\end{center}
\end{figure}

In the previous section, we exploited the vanishing of the atomic
decay rate in order to determine the optimal input pulse $\ket{\text{opt}_N}$
that implies a dramatic suppression of atomic decay rate.
For an appropriately large cavity length, moreover, this optimal
pulse leads to an atom-field evolution with $g^\circ_N >
\kappa^\circ_N \gg \gamma^\circ_N(\omega_a)$. Although the parameter
window for $N>1$, in which $\gamma^\circ_N(\omega_a)$ becomes
efficiently suppressed, is rather small to be accessible in
practice, the results we obtained provide us with the relevant
insights about the cavity-QED like behavior of the coupled
atom-cavity-pulse system shown in Fig.~\ref{fig2}(a).

Apparently, the lateral profile (\ref{opt}) has a complicated shape that
makes the experimental generation of the respective spatial profile
$\varphi^\text{opt}_\alpha(x,y, f, N)$ very challenging. This conclusion
along with a small parameter window for $N > 1$, in which the atomic
decay becomes suppressed, suggest us to consider a specific input
pulse that can be easily tailored in an experiment. In this section, we
consider the Hermite-Gaussian beams TEM$_{1,0}$ and TEM$_{0,1}$ of
the waist $w$, which we identify with the $\parallel$ and $\bot$
polarization-components of $\varphi_\alpha^G (x,y,w)$, respectively.
Using these beams, we analyze the cavity-QED parameters as functions
of $w$ and the number of quasi-modes coupled to the atom, $N$.

We observe that the dependence in (\ref{opt}) on the radial part of $\kt$
[see Fig.~\ref{fig1}(b)] poses the main difficulty concerning the generation
of this lateral profile in practice. The dependence on the angular part of $\kt$,
in contrast, is simple and is encoded in the atom-field coupling (\ref{lambda})
\begin{equation}
\lambda_{\parallel, n}(\kt) = \lambda_{\parallel, n}^\circ(k) \,
\cos \vartheta \, ; \quad \lambda_{\bot, n}(\kt) = \lambda_{\bot,
n}^\circ(k) \, \sin \vartheta \, , \notag
\end{equation}
where the dipole orientation assumption $\vecP \cdot \hz = 0$ and
the explicit form of mode functions (\ref{u-modes}) have been used.
Motivated by this simple angular dependence (preserved by the
Fourier transform), we suggest the identification
\begin{subequations}\label{hg-beams}
\begin{eqnarray}
\varphi_\parallel^G (x,y,w) &=& \frac{1}{\sqrt{2}} \, \mathcal{F}_1(x,w) \,
                                \mathcal{F}_0(y,w) \quad [\text{TEM}_{1,0}] \, ; \\
\varphi_\bot^G (x,y,w) &=& \frac{1}{\sqrt{2}} \, \mathcal{F}_0(x,w) \,
                           \mathcal{F}_1(y,w) \quad [\text{TEM}_{0,1}] \, , \qquad
\end{eqnarray}
\end{subequations}
where
\begin{equation}
\mathcal{F}_n(z,w) = \left( \frac{2}{\pi} \right)^{\frac{1}{4}}
   \sqrt{\frac{1}{2^{n + \frac{1}{2}} \, n! \, w}} \, H_n\left( \frac{z}{w} \right)
   e^{- \frac{z^2}{2 w^2}} \, . \quad
\end{equation}
\begin{figure*}[!ht]
\begin{center}
\includegraphics[width=0.975\textwidth]{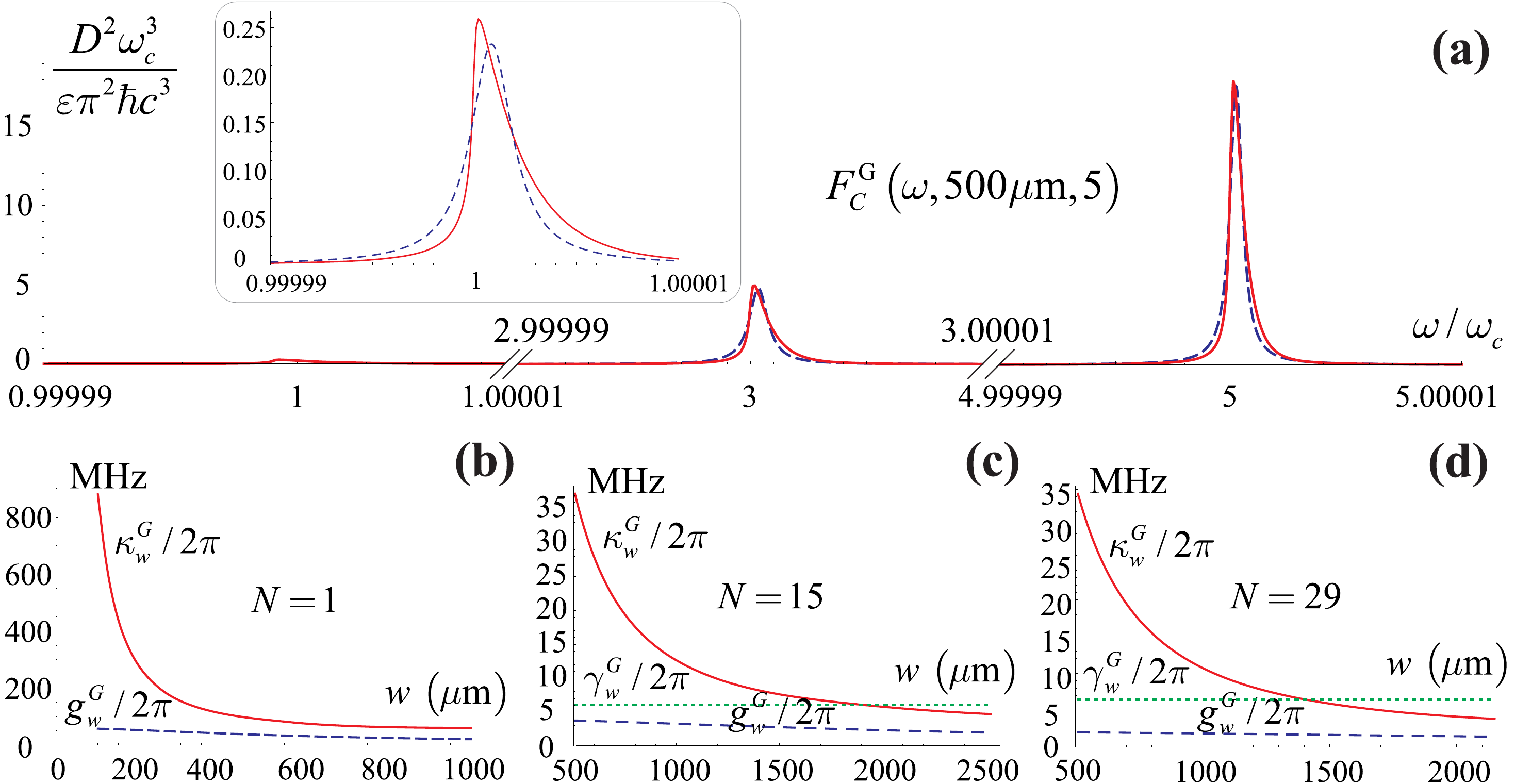} \\
\caption{(Color online) (a) The cavity form-factor (solid curve) and the associated
Lorentzian (dashed curve) for $N=5$ and $w = 500 \mu$m. The last peak resembles an
almost ideal Lorentzian. (b), (c), and (d) Cavity-QED parameters $g^G_w$ (dashed
curve), $\kappa^G_w$ (solid curve), and $\gamma^G_w$ (dotted curve) as functions
of $w$ for $N=1$, $15$, and $29$, respectively. See text for details.}
\label{fig6}
\end{center}
\end{figure*}

The lateral profiles (\ref{hg-beams}) are the simplest and
experimentally most feasible beams which exhibit the same angular
dependence in physical space as the Fourier transform of
$\widetilde{\varphi}^\text{opt}_\alpha(\kt, f, N)$. These beams
depend on the waist $w$ and fulfill the normalization condition
(\ref{norm0}). The corresponding input pulse, therefore, takes the
form
\begin{equation}\label{hg-pulse}
\ket{HG} = \sum_\alpha \intd \kv \, \widetilde{\psi}^\text{opt}_{\omega_\circ}(\kz) \,
                             \widetilde{\varphi}^G_\alpha(\kt, w) \,
                             b^\dag_\alpha(\kv) \ket{\text{vac}} \, , 
\end{equation}
where $\widetilde{\psi}^\text{opt}_{\omega_\circ}$ was defined
in the previous subsection, while
\begin{equation}\label{hg-beams1}
\widetilde{\varphi}^G_\alpha(\kt, w) = \frac{w^2 k}{\im \sqrt{\pi}} \,
          e^{- \frac{w^2 k^2}{2}} \left( \cos \left[ \vartheta \right] \delta_{\alpha, \parallel}
                                    + \sin \left[ \vartheta \right] \delta_{\alpha, \bot} \right)
\end{equation}
satisfies the normalization condition (\ref{norm1}). We insert this
lateral profile into Eq.~(\ref{ffc}) and obtain
\begin{equation}\label{ffc-hg0}
F^G_{C}(\omega, w, N) =
   \frac{1}{c} \left| \sum_{n, \alpha}^{N} \intd \kt \, \lambda_{\alpha, n}(\kt) \, 
   \widetilde{\varphi}^G_\alpha(\kt, w) \, A_n \left( \omega, k \right) \right|^2 , \qquad
\end{equation}
which becomes after the evaluation
\begin{eqnarray}\label{ffc-hg}
F^G_{C}(\omega, w, N) =
  && \frac{\mathcal{D}^2 \, \omega_c^3}{\epsilon \, \pi^2 \hbar \, c^3}
      \left( \frac{\tau^2 \, w^4 \, \omega_c^4}{4 \, \pi^2 \, c^4} \right) \times \notag \\
  && \hspace{1cm} \left| F^\bullet_C \left( \frac{\omega}{\omega_c}, \frac{w \, \omega_c}{c}, N \right) \right|^2 ,
\end{eqnarray}
with the notation
\begin{eqnarray}
&& F^\bullet_{C}(u, v, N) = \sum_n^N \sin \left[ \frac{3 \pi n}{2} \right] \times \notag \\
&& \qquad \int_n^\infty \frac{\sqrt{s(s^2 - n^2)} (n + s) \, e^{- \frac{v^2}{2} (s^2 -n^2)} \, ds}
                      {u - s - \im \tau^2 / (4 \, \pi)} \, .
\end{eqnarray}

We recall that only in the case when the cavity form-factor
resembles a sharply peaked function, it plays the role of spectral 
mode density in a cavity-QED system with losses. We have checked 
that, in contrast to the optimal cavity form-factor (\ref{ffc-opt}), 
the form-factor (\ref{ffc-hg}) yields only deformed Lorentzians for
small beam waists $w$. However, the larger the waist we consider,
the less deformed the peaks we obtain. Considering the resonator
that accommodates only one atomic wavelength, for instance, in
Figs.~\ref{fig5}(a) and (b) we display $F^G_{C}(\omega, w, 1)$
(solid curves) for $w = 100 \mu$m and $500 \mu$m, respectively,
where the condition $\omega_c = \omega_a$ along with the atomic data
(\ref{data}) have been used. The dashed curves depict the
Lorentzians obtained as the best fit to the respective solid curves.
It is seen that the solid curve in Fig.~\ref{fig5}(a) is notably
deformed with regard to the dashed one, while both curves in
Fig.~\ref{fig5}(b) almost coincide.

We have checked, furthermore, that the total form-factor gives the
major contribution to the atomic decay rate $\gamma^G_w(\omega) =
F_{T}(\omega, 1) - F^G_{C}(\omega, w, 1)$. This leads to an
efficient suppression of the atomic decay only in the region $\omega
< \omega_c$ [see Fig.~\ref{fig2}(c)]. Using
the Lorentzians (dashed curves) from Figs.~\ref{fig5}(a) and (b), we
calculate the cavity-QED parameters
\begin{subequations}
\begin{eqnarray}
&& \left( g^G_w, \kappa^G_w, \gamma^G_w(0.99 \, \omega_a) \right) =
        2 \pi \left( 57, 879, 0.5 \; 10^{-4} \right) \text{MHz} , \qquad \label{params3} \\
&& \left( g^G_w, \kappa^G_w, \gamma^G_w(0.99 \, \omega_a) \right) =
        2 \pi \left( 35, 89, 0.5 \; 10^{-4} \right) \text{MHz} ,  \label{params4}
\end{eqnarray}
\end{subequations}
for the beam waists $w = 100 \mu$m and $500 \mu$m, respectively, 
where the central frequency of input pulse is slightly detuned from 
the atomic transition frequency, that is $\omega_\circ = 0.99 \, \omega_a$.
These parameters suggest that the cavity relaxation rate drops for a
larger waist of the beam, while the atomic decay rate remains
negligible if compared to other parameters. We stress, however, that
a large waist $w$ of the input beam requires a large surface size
of the resonator which, from an experimental point of view, is
likely incompatible with a small-volume resonator that accommodates 
just one atomic wavelength.

Before we turn to a big-volume resonator that accommodates $N$
atomic wavelengths, we remark that, in contrast to the optimal
form-factor (\ref{ffc-opt}) that produces a single peak at $\omega =
f$, the form-factor (\ref{ffc-hg}) produces a series of peaks at
$\omega = n \, \omega_c$ ($n=1,3,5, \ldots$), which resemble nice
Lorentzians only for large waists $w$. To illustrate this feature,
we consider the resonator that accommodates five atomic wavelengths,
that is $\omega_a = 5 \, \omega_c$ and $N=5$. In Fig.~\ref{fig6}(a),
we display $F^G_{C}(\omega, 500 \, \mu \text{m}, 5)$ (solid curves)
using the atomic data (\ref{data}). As in the previous Figure, the
dashed curves depict the Lorentzians obtained as the best fit to the
respective solid curves.

It can be seen that the cavity form-factor produces three different
peaks at $\omega / \omega_c = 1,3,$ and $5$, while only the peak at
$\omega = \omega_a$ resembles an almost perfect Lorentzian. The three
Lorentzians (dashed curves) in Fig.~\ref{fig6}(a) yield
\begin{subequations}\label{params5-7}
\begin{eqnarray}
&& \left( g^G_w, \kappa^G_w, \gamma^G_w(0.99 \, \omega_c) \right) =
        2 \pi \left( 2, 175, 0.4 \; 10^{-5} \right) \text{MHz} \, , \qquad \label{params5} \\
&& \left( g^G_w, \kappa^G_w, \gamma^G_w(3.99 \, \omega_c) \right) =
        2 \pi \left( 6, 68, 0.8 \right) \text{MHz} \, ,  \label{params6} \\
&& \left( g^G_w, \kappa^G_w, \gamma^G_w(4.99 \, \omega_c) \right) =
        2 \pi \left( 10, 45, 4.8 \right) \text{MHz} \, ,  \label{params7}
\end{eqnarray}
\end{subequations}
corresponding to central frequencies of the input pulse which are 
slightly detuned from $\omega_c$, $3 \, \omega_c$, and $5 \, \omega_c$,
respectively. We see that the last peak resembles not only an almost
perfect Lorentzian, but also implies a higher $g^G_w$ and a smaller
cavity relaxation rate $\kappa^G_w$ than those parameters associated
with the other two peaks. The atomic decay rate $\gamma^G_w$, in
contrast, increases due to the major contribution of the total
form-factor that vanishes in the region $\omega < \omega_c$.

After we pointed out the main features of the cavity form-factor
(\ref{ffc-hg}) for $N=1$ and $N=5$, let us consider a resonator that
accommodates $N = 1$, $15$, and $29$ atomic wavelengths and attempt
to reveal the dependence of cavity-QED parameters on the beam waist
$w$. Although we noticed that a large waist of the input beam is
likely incompatible with a small cavity size, for completeness, we
include the case $N=1$ in our considerations. From the case $N=5$
analyzed above, we learned that the form-factor produces $(N+1)/2$
peaks, such that the last peak (that matches the atomic transitions
frequency) resembles the most perfect Lorentzian. Motivated by this
essential requirement (that justifies our approach), we calculate
below the cavity-QED parameters associated with this last peak,
where the central frequency of pulse is slightly detuned from the 
atomic transition frequency.

In Figs.~\ref{fig6}(b)-(d), we display $g^G_w / 2 \, \pi$ (dashed
curve), $\kappa^G_w / 2 \, \pi$ (solid curve), and $\gamma^G_w / 2
\, \pi$ (dotted curve) as functions of $w$ for the above mentioned
three values of $N$. Although in all three figures the cavity
relaxation rate is efficiently suppressed for large $w$, it still
remains notably higher than the atom-field coupling strength
$g^G_w$. The atomic decay rate that is negligibly small for $N=1$
oversteps slightly $g^G_w$ for $N>1$, which is in agreement with the
observations we already made [see (\ref{params5-7})]. It is clearly
seen, furthermore, that the atom-field evolution for $N=1$ implies
$\kappa^G_w > g^G_w \gg \gamma^G_w(0.99 \, \omega_a)$. For $N>1$ and
small $w$, in contrast, the atom-field evolution implies $\kappa^G_w
> \gamma^G_w(0.99 \, \omega_a) > g^G_w$, while for a reasonably high
$w$ the same evolution implies $\gamma^G_w(0.99 \, \omega_a) >
\kappa^G_w > g^G_w$.

To summarize this section, we considered the Hermite-Gaussian input
pulse (\ref{hg-pulse}) instead of the optimal pulse $\ket{\text{opt}_N}$. 
Using the cavity form-factor (\ref{ffc-hg}) associated with this
(experimentally feasible) input pulse, we studied the dependence of
cavity-QED parameters on the beam waist and the number of cavity
quasi-modes coupled to an atom. In contrast to the results we
obtained in the previous section, the atomic decay rate becomes
dramatically suppressed only for $N=1$ and the central frequency 
that is slightly detuned from the atomic transition 
frequency. For $N>1$, however, the atomic decay rate becomes 
non-negligible and it oversteps the atom-field coupling strength, 
while for a reasonably large waist of the beam, the atomic decay 
rate oversteps both the atom-field coupling  strength and the cavity 
relaxation rate. We conclude, therefore, that an input beam that 
reproduces only the angular part of the optimal lateral profile 
$\widetilde{\varphi}^\text{opt}_\alpha$, is insufficient to achieve
the cavity-QED evolution, such that the atom-field energy exchange
develops faster than the losses due to the cavity relaxation and the
atomic decay.

\section{Summary and Outlook}

In this paper, we generalized the framework of Ref.~\cite{K} by
means of (i) an \textit{ab-initio} derivation of the
atom-cavity-pulse Hamiltonian, (ii) including the
$\parallel$-component of light polarization, (iii) treating the
cavity relaxation as a function of transmissivity and reflectivity,
(iv) considering a realistic (non-averaged) atomic dipole, and by
(v) analyzing the resonators which accommodate $N \geq 1$ atomic
wavelengths. Using this generalized framework, we derived the
cavity-QED parameters and revealed their dependence on the atom and
cavity characteristics, number of cavity quasi-modes coupled to the
atom, and the spatial distribution of the input pulse. The optimal
spatial distribution that yields vanishing of the atomic decay rate
was determined. We calculated cavity-QED parameters for this optimal
distribution and found that the atomic decay is efficiently
suppressed once this optimal pulse with a proper frequency distribution
penetrates the resonator. We demonstrated that the suppression of
atomic decay occurs even for a central frequency that is larger than 
the cut-off frequency in a larger resonator.

Besides this optimal pulse, the scenario in which a Hermite-Gaussian 
beam penetrates the resonator was considered. We discussed in detail 
this scenario and revealed the dependence of the cavity-QED parameters 
on the beam waist and the cavity length. In contrast to the results 
obtained for an optimal pulse, the atomic decay becomes suppressed 
only in a resonator that accommodates one single atomic wavelength.
We concluded that an input pulse that reproduces only the angular part 
of the optimal spatial distribution is insufficient and so, also the 
radial profile has to resemble the respective profile associated with 
the optimal pulse.

We found that the spatial distribution of the
input pulse determine the radiative properties of an atom coupled to
a planar resonator. By providing the coupled atom-cavity system
with an input pulse that resembles the optimal pulse or the spatial
distribution that maximizes the non-cavity form-factor, therefore,
one can either suppress completely the atomic decay or enhance the
spontaneous emission on demand. This property suggests that our
system can act as a quantum memory for long-term storage of a single
qubit, where the two-level atom inserted into the resonator is
interpreted as a qubit, while the (controlled) atomic decay
constitutes the main source of dephasing and decoherence. Besides
the storage of a qubit, a quantum memory should also provide
reliable write-in and read-out mechanisms, which together with a
quantitative characterization of the memory itself shall be
addressed in our future works.

To conclude, we showed that our atom-cavity-pulse system can behave
as a cavity-QED system exhibiting the spectral mode density of a
completely (3D) confined system with losses. We remark, however,
that our system is a typical 1D confined system, in which only one
component of the photon field is confined, while the two remaining
components propagate in free space. On the other hand, although an
efficient and deterministic atom-light coupling in free space poses
a serious experimental challenge \cite{apb89, jmo58, np7}, an
atom-light interface in free space may open a route towards scalable
quantum networking due to a moderate demand of physical resources.
The remark above suggests that the studied atom-cavity-pulse system 
can be interpreted as a system that combines both cavity-QED and free 
space features. Indeed, by making the mirrors of the planar resonator
completely transparent, we would (effectively) reproduce the
interaction of an atom and an input pulse in free space. We stress
that, although justification of the results obtained in this paper
relies on the restriction $\tau \ll 1$ [see (\ref{new})], in
principle, this restriction can be reasonably relaxed at the expense
of introducing nonorthogonal modes in our framework \cite{pra62}.

\begin{acknowledgments}

We thank the BMBF for support through the QuOReP program. We also
thank Gernot Alber for helpful comments and suggestions.

\end{acknowledgments}

\end{document}